\documentclass[lettersize,journal]{IEEEtran}
\usepackage{amsmath,amsfonts}
\usepackage{algorithmic}
\usepackage{algorithm}
\usepackage{array}
\usepackage[caption=false,font=normalsize,labelfont=sf,textfont=sf]{subfig}
\usepackage{textcomp}
\usepackage{stfloats}
\usepackage{url}
\usepackage{verbatim}
\usepackage{graphicx}
\usepackage{cite}
\usepackage{amssymb}
\usepackage[table, xcdraw]{xcolor} 
\usepackage[hidelinks]{hyperref}
\newcommand{\refappendix}[1]{\hyperref[#1]{Appendix~\ref*{#1}}} 
\newcommand*{\toolbox}{\fontfamily{pcr}\selectfont} 
\DeclareMathOperator*{\argmin}{argmin} 
\usepackage{float}
\usepackage{arydshln} 
\usepackage{booktabs}  

\usepackage{listings}

\definecolor{mygreen}{rgb}{0,0.6,0}
\definecolor{mygray}{rgb}{0.5,0.5,0.5}
\definecolor{mymauve}{rgb}{0.58,0,0.82}

\lstdefinelanguage{XMLfastr}
{
	morestring=[b]",
	morestring=[s]{>}{<},
	morecomment=[s]{<?}{?>},
	stringstyle=\color{black},
	identifierstyle=\color{darkblue},
	keywordstyle=\color{cyan},
	morekeywords={authors, targets}
}

\begin{document}
	
		\title{An automated machine learning framework to optimize radiomics model construction validated on twelve clinical applications}
		
		\author{Martijn~P.A.~Starmans, Sebastian~R.~van~der~Voort, Thomas~Phil, Milea~J.M.~Timbergen, Melissa~Vos, Guillaume~A.~Padmos, Wouter~Kessels, David~Hanff, Dirk~J.~Gr\"unhagen, Cornelis~Verhoef, Stefan~Sleijfer, Martin~J.~van~den~Bent, Marion~Smits, Roy~S.~Dwarkasing, Christopher~J.~Els, Federico~Fiduzi, Geert~J.L.H.~van~Leenders, Anela~Blazevic, Johannes~Hofland, Tessa~Brabander, Renza~A.H.~van~Gils, Gaston~J.H.~Franssen, Richard~A.~Feelders, Wouter~W.~de~Herder, Florian~E.~Buisman, Francois~E.J.A.~Willemssen, Bas~Groot~Koerkamp, Lindsay~Angus, Astrid~A.M.~van~der~Veldt, Ana~Rajicic, Arlette~E.~Odink, Mitchell~Deen, Jose~M.~Castillo~T., Jifke~Veenland, Ivo~Schoots, Michel~Renckens, Michail~Doukas, Rob~A.~de~Man, Jan~N.M.~IJzermans, Razvan~L.~Miclea, Peter~B.~Vermeulen, Esther~E.~Bron, Maarten~G.~Thomeer, Jacob~J.~Visser, Wiro~J.~Niessen, Stefan~Klein, for the Alzheimer’s Disease Neuroimaging Initiative
			
			\thanks{M.P.A.~Starmans, S.R.~van~der~Voort, T.~Phil, M.J.M.~Timbergen, M.~Vos, G.~A.~Padmos, W.~Kessels, D.~Hanff, D.J.~Gr\"unhagen, C.~Verhoef, S.~Sleijfer, M.J.~van~den~Bent ,M.~Smits, R.S.~Dwarkasing, C.J.~Els, F.~Fiduzi, G.J.L.H.~van~Leenders, A.Blazevic, J.Hofland, T.Brabander, R.A.H.~van~Gils, G.J.H.~Franssen, R.~A.~Feelders, W.W.~de~Herder, F.~E.~Buisman, F.E.J.A.~Willemssen, B.~Groot~Koerkamp, L.~Angus, A.A.M.~van~der~Veldt, A.~Rajicic, A.E.~Odink, M.~Deen, J.M.~Castillo~T., J.~Veenland, I.~Schoots, M.~Renckens, M.~Doukas, R.A.~de~Man, J.N.M.~IJzermans, E.E.~Bron, M.G.~Thomeer, J.J.~Visser, W.J.Niessen and S. Klein are with the Departments of Radiology and Nuclear Medicine, Oral and Maxillofacial Surgery, Medical Oncology, Surgical Oncology, Neurology, Pathology, Internal Medicine, Surgery, and Hepatology, Erasmus MC, Rotterdam, the Netherlands. Correspondence: m.starmans@erasmusmc.nl}
			\thanks{R.L.~Miclea is with the Department of Radiology and Nuclear Medicine, Maastricht UMC+, Maastricht, the Netherlands}
			\thanks{P.B.~Vermeulen is with the Translational Cancer Research Unit, Department of Oncological Research, Oncology Center, GZA Hospitals Campus Sint-Augustinus and University of Antwerp, Antwerp, Belgium}
			\thanks{W.J.~Niessen is also with the Faculty of Applied Sciences, Delft University of Technology, Delft, the Netherlands}
			\thanks{Data used in preparation of this article were obtained from the Alzheimer’s Disease Neuroimaging Initiative (ADNI) database (\url{adni.loni.usc.edu}). As such, the investigators within the ADNI contributed to the design and implementation of ADNI and/or provided data but did not participate in analysis or writing of this report. A complete listing of ADNI investigators can be found at: \url{https://adni.loni.usc.edu/wp-content/uploads/how_to_apply/ADNI_Acknowledgement_List.pdf}}
	}
	
	\markboth{}%
	{Starmans \MakeLowercase{\textit{et al.}}: An automated machine learning framework to optimize radiomics model construction validated on twelve clinical applications}
	
	
	\maketitle
	
	\begin{abstract}
		Predicting clinical outcomes from medical images using quantitative features (“radiomics”) requires many method design choices, Currently, in new clinical applications, finding the optimal radiomics method out of the wide range of methods relies on a manual, heuristic trial-and-error process. We introduce a novel automated framework that optimizes radiomics workflow construction per application by standardizing the radiomics workflow in modular components, including a large collection of algorithms for each component, and formulating a combined algorithm selection and hyperparameter optimization problem. To solve it, we employ automated machine learning through two strategies (random search and Bayesian optimization) and three ensembling approaches. Results show that a medium-sized random search and straight-forward ensembling perform similar to more advanced methods while being more efficient. Validated across twelve clinical applications, our approach outperforms both a radiomics baseline and human experts. Concluding, our framework improves and streamlines radiomics research by fully automatically optimizing radiomics workflow construction. To facilitate reproducibility, we publicly release six datasets, software of the method, and code to reproduce this study.
	\end{abstract}
	
	
	\section{Introduction}
	\label{sec:introduction}
	There has been a paradigm shift in health care, moving from a reactive, one-size-fits-all approach, towards a more proactive, personalized approach \cite{RN925, RN928, RN929}. To aid in this process, personalized medicine generally involves clinical decision support systems such as \textit{biomarkers}, which relate specific patient characteristics to some biological state, outcome, or condition. To develop such biomarkers, medical imaging has gained an increasingly important role \cite{RN925, RN1082, RN326}. Specifically, machine learning, both using conventional and deep learning methods, has shown to be highly successful for medical image classification and has thus become the \textit{de facto} standard. Within the field of radiology, ``radiomics'' has been coined to describe the use of a large number of quantitative medical imaging features in combination with (typically conventional, non-deep) machine learning to create biomarkers \cite{RN130}. These for example relate to diagnosis, prognosis, histology, treatment and surgery planning or response, drug usage, and genetic mutations. The rise in popularity of radiomics has resulted in a large number of papers and a wide variety of methods \cite{Li2022, Bera2022, RN261, RN691, RN436, RN493, RN674, RN798, RN799, RN35, RN686, RN688, RN800, RN1082, RN1079, RN1227, RN447, RN505, RN1089}: as of March 3, 2025, a PubMed search using the term ``radiomics'' yielded 16,015 publications.
	
	Currently, in a new clinical application, finding the optimal radiomics method out of the wide range of available options has to be done manually through a heuristic trial-and-error process. This process has several disadvantages, as it: 1) is time-consuming; 2) requires expert knowledge; 3) does not guarantee that an optimal solution is found; 4) negatively affects  reproducibility \cite{Li2022, Bera2022, RN686, RN436, RN674}; 5) has a high risk of overfitting when not carefully conducted \cite{RN684, RN686}; and 6) limits translation to clinical practice \cite{RN674}. While the lack of method standardization has been identified as a major challenge for radiomics \cite{Li2022, Bera2022}, we observed that most published radiomics methods roughly consist of the same steps: image segmentation, preprocessing, feature extraction, feature selection, and classification. Hence, we hypothesized that it should be possible to automatically find the optimal radiomics model in a new clinical application by collecting numerous methods in one framework and systematically comparing and combining all included components.
	
	The aim of this study is to streamline radiomics research, facilitate radiomics' reproducibility, and ultimately simplify radiomics use in (new) clinical applications. To this end, we propose an end-to-end (i.e., images to biomarker) framework to fully automatically optimize radiomics workflow construction per application by exploiting recent advances in automated machine learning (AutoML) \cite{RN694}. We define a radiomics workflow as a specific combination of algorithms and their associated hyperparameters, i.e., parameters that need to be set before the learning step. We focus on conventional radiomics pipelines, i.e., using non-deep machine learning, as: 1) conventional methods are quick to train, hence AutoML is feasible to apply; 2) the search space is relatively clear as conventional workflows typically follow the same steps, further enhancing AutoML feasibility; 3) as there is a large number of conventional radiomics papers, the impact of such a method is potentially large; and 4) conventional radiomics is also suitable for small datasets, which is relevant for rare diseases \cite{RN35, RN686}. We coin our proposed novel framework {\toolbox{WORC}} (Workflow for Optimal Radiomics Classification).
	
	The key contributions of this work can be summarized as follows:
	\begin{enumerate}
		\item We propose a modular radiomics search space with standardized components (i.e., fixed inputs, functionality, and outputs) to facilitate the use of AutoML. To create a comprehensive search space, we include many commonly used algorithms and their associated hyperparameters for each component.
		\item We propose the construction of a radiomics workflow per application as a Combined Algorithm Selection and Hyperparameter (CASH) optimization problem \cite{RN56}, including both the choice of algorithms and their associated hyperparameters. To this end, we reformulate and extend the original CASH problem to the complete radiomics workflow. 
		\item  We compare various approaches for solving the CASH problem, including two search strategies: an efficient random search and Bayesian optimization, the state-of-the-art in AutoML \cite{RN694, RN1358}. To exploit prior knowledge, we introduce a light fingerprinting mechanism to reduce the search space based on the application at hand. Additionally, we propose and compare various ensembling approaches to boost performance and stability. 
		\item We extensively evaluate our framework by conducting a large number of experiments on twelve different clinical applications using three publicly available datasets and nine in-house datasets and compare our framework to a radiomics baseline.
		\item We publicly release six datasets with data of in total 930 patients to facilitate reproducibility \cite{WORCDatabase}, inviting the machine learning community to evaluate novel methods on these radiomics benchmark datasets. Additionally, we have made the software implementation of our framework, and the code to perform our experiments on all datasets open-source \cite{RN63, WORCDatabasesoftware}.
	\end{enumerate}
	
	\subsection{Background and related work} \label{sec: background}
	To outline this study's context, we present some background on typical radiomics studies. Generally, a radiomics study can be seen as a collection of various steps: data acquisition and preparation, segmentation, feature extraction, and data mining \cite{RN493}. In this study, we consider the data, i.e., the images, ground truth labels, and segmentations, to be given; data acquisition and segmentation algorithms are out of scope. 
	
	First, radiomics workflows commonly start by preprocessing the images and segmentations to compensate for undesired data variations \cite{RN436}. Examples are image intensity normalization, or resampling to the same voxel spacing.
	
	Second, quantitative image features are computationally extracted. As most radiomics applications are in oncology, feature extraction algorithms generally focus on describing properties of a specific region of interest, e.g., tumors, and require a segmentation. Features are typically split in three groups \cite{RN238, RN539}: 1) first-order or histogram, quantifying intensity distributions; 2) morphology, quantifying shape; and 3) higher-order or texture. Many open-source toolboxes for radiomics feature extraction exist \cite{RN686}.
	
	Lastly, the data mining component may itself consist of a combination of various components: 1) feature imputation; 2) feature scaling; 3) feature selection; 4) dimensionality reduction; 5) resampling; 6) (machine learning) algorithms to find relationships between the remaining features and certain outcomes. While data mining is often seen as one single component, we explicitly split this step into separate components.
	
	AutoML has previously been used in radiomics using Tree-based Pipeline Optimization Tool (TPOT) \cite{Olson2019} to predict the H3 K27M mutation in midline glioma \cite{RN491}, to predict invasive placentation \cite{RN1114}, and in a study using the {\toolbox WORC} database we publish as part of the current paper \cite{RN1107}. In this study, we optimized construction of the complete radiomics workflow from images to prediction, included a large collection of commonly used radiomics algorithms and parameters in the search space, compared various AutoML and ensembling approaches, and extensively evaluated our approach and its generalizability in twelve different applications. 
	
	We have previously published clinical studies in which our {\toolbox WORC} framework has been applied to develop and evaluate dedicated biomarkers (e.g., \cite{RN16, RN480, RN574, CRLMPaper, RN678, starmans2024BLT, RN791, RN923, RN868, RN1247, RN1246}). These studies focused on the evaluation of the clinical relevance and impact of the resulting biomarkers and only describe the application of (previous versions of) {\toolbox WORC}. In the current study, we present the complete method itself in detail; introduce a light fingerprinting mechanism; simultaneously evaluate the framework on multiple clinical applications using a single harmonized configuration; analyze in-depth the influence of random search, Bayesian optimization, and various ensemble settings on performance and stability; and compare the performances to a radiomics baseline.
	
	\section{Methods} \label{sec: methods}
	This study focuses on binary classification problems, as these are most common in radiomics \cite{RN686}. In the {\toolbox WORC} toolbox, we have straight-forward extended our CASH reformulation to multiclass, multilabel, and regression problems, and have included the related search spaces, which are described in the {\toolbox{WORC}} documentation \cite{RN540}. 
	
	\subsection{Automatic workflow optimization} \label{sec: adaptiveworkflowoptimization}
	The aim of our framework is to automatically optimize radiomics workflow construction out of a large number of algorithms and associated hyperparameters. To this end, we identified three key requirements. First, our optimization strategy should adapt the workflow per application as the optimal combination may vary. Second, while model selection is typically performed before hyperparameter tuning, it has been shown that these two problems are not independent \cite{RN56}. Thus, combined optimization is required. Third, to prevent over-fitting, all optimization should be performed on a training dataset, independent from the test dataset \cite{RN694, RN684, RN686}. As manual model selection and hyperparameter tuning is not feasible in a large search space and not reproducible, all optimization should be automatic.
	
	\subsubsection{The Combined Algorithm Selection and Hyperparameter optimization problem}
	To address the three identified key requirements, we propose to formulate the complete radiomics workflow construction as a CASH optimization problem, adopting the previous CASH definition in AutoML for machine learning model optimization \cite{RN56}. For a single algorithm, the associated hyperparameter space consists of all possible values of all the associated hyperparameters. In machine learning, given a dataset $\mathcal{D} = \left\{(\vec{x_1}, y_1), \ldots, (\vec{x_n}, y_n)\right\}$ consisting of features $\vec{x}$ and ground truth labels $y$ for $n$ objects or samples, and a set of algorithms $\mathcal{A} = \left\lbrace A^{(1)}, \ldots, A^{(m)} \right\rbrace$ with associated hyperparameter spaces $\boldsymbol{\Lambda}^{(1)}, \ldots, \boldsymbol{\Lambda}^{(m)}$ consisting of hyperparameters $\boldsymbol{\lambda}^{(m)} = \lambda_1, \ldots, \lambda_l \in \boldsymbol{\Lambda}^{(m)}$, the CASH optimization problem is to find the algorithm $A^{*}$ and associated hyperparameter set $\boldsymbol{\lambda}^{*}$ that minimize the loss $\mathcal{L}$:
	
	\begin{equation}
		A^{*}, \boldsymbol{\lambda}^{*} \in \argmin_{A^{(j)} \in \mathcal{A}, \boldsymbol{\lambda} \in \boldsymbol{\Lambda}^{(j)}} \frac{1}{k_{\text{training}}} \sum_{i=1}^{k_{\text{training}}} \mathcal{L} \left(A^{(j)}_{\boldsymbol{\lambda}}, \mathcal{D}^{(i)}_{\text{train}}, \mathcal{D}^{(i)}_{\text{valid}} \right),
		\label{eq: CASH}
	\end{equation}
	where a cross-validation with $k_{\text{training}}$ iterations is used to define subsets of the full dataset for training ($\mathcal{D}^{(i)}_{\text{train}}$) and validation ($\mathcal{D}^{(i)}_{\text{valid}}$). In order to combine model selection and hyperparameter optimization, the problem can be reformulated as a pure hyperparameter optimization problem by introducing a new hyperparameter $\lambda_r$ that selects between algorithms: $\boldsymbol{\Lambda} = \boldsymbol{\Lambda}^{(1)} \bigcup \ldots \bigcup \boldsymbol{\Lambda}^{(m)} \bigcup \{\lambda_r \}$ \cite{RN56}. Thus, $\lambda_r$ defines which algorithm from $\mathcal{A}$ and which associated hyperparameter space $\boldsymbol{\Lambda}$ are used. This results in:
	
	\begin{equation}
		\boldsymbol{\lambda}^{*} \in \argmin_{\boldsymbol{\lambda} \in \boldsymbol{\Lambda}} \frac{1}{k_{\text{training}}} \sum_{i=1}^{k_{\text{training}}} \mathcal{L} \left( \boldsymbol{\lambda}, \mathcal{D}^{(i)}_{\text{train}}, \mathcal{D}^{(i)}_{\text{valid}} \right).
		\label{eq: CASH2}
	\end{equation}
	
	We extend the CASH optimization problem to the complete radiomics workflow, consisting of various components. The parameters of all algorithms are treated as hyperparameters. Furthermore, instead of introducing a single hyperparameter to select between algorithms, we define multiple algorithm selection hyperparameters. Two categories are distinguished: 1) for optional components,  \textit{activator} hyperparameters are introduced to determine whether the component is used; and 2) for mandatory components, integer \textit{selector} hyperparameters are introduced to select one of the available algorithms. Optional components that contain multiple algorithms have both \textit{activator} and \textit{selector} hyperparameters. We thus reformulate CASH optimization for a collection of $t$ algorithm sets $\mathcal{A_C} = \mathcal{A}_1 \bigcup \ldots \bigcup \mathcal{A}_t$ and the collection of associated hyperparameter spaces $\boldsymbol{\Lambda}_C = \boldsymbol{\Lambda}_1 \bigcup \ldots \bigcup \boldsymbol{\Lambda}_t $. Including \textit{activator} and \textit{selector} model selection parameters within the collections, similar to \autoref{eq: CASH2}, this results in:
	
	\begin{equation}
		\boldsymbol{\lambda}^{*} \in \argmin_{\boldsymbol{\lambda_C} \in \boldsymbol{\Lambda}_C} \frac{1}{k_{\text{training}}} \sum_{i=1}^{k_{\text{training}}} \mathcal{L} \left( \boldsymbol{\lambda}_{C}, \mathcal{D}^{(i)}_{\text{train}}, \mathcal{D}^{(i)}_{\text{valid}} \right).
		\label{eq: CASHRadiomics}
	\end{equation}
	
\begin{figure*}
		\centering
		\includegraphics[width=\textwidth]{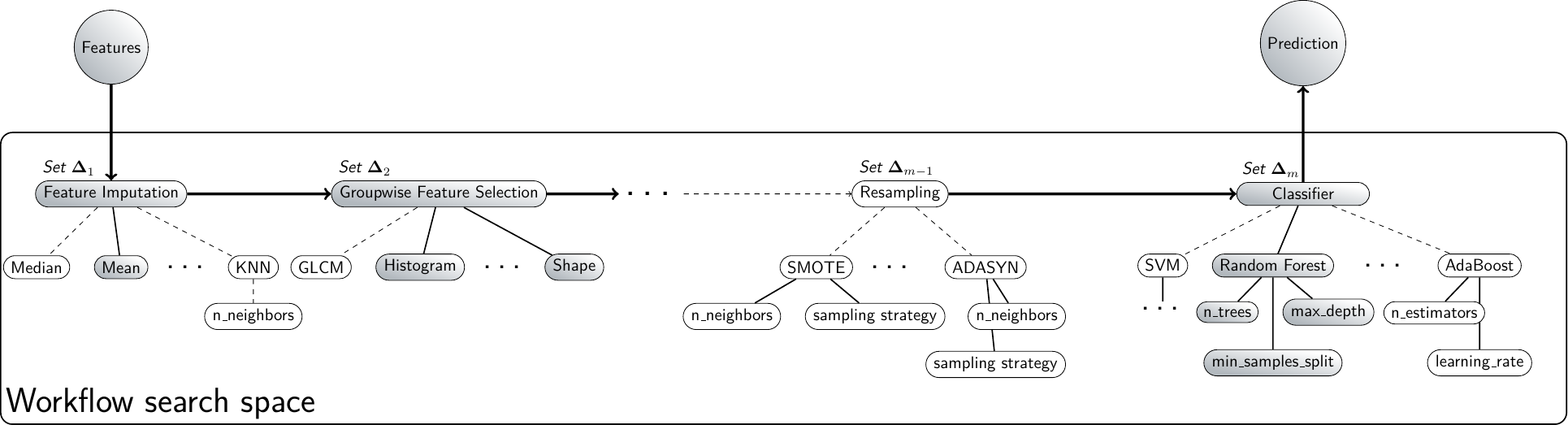}
		\caption{Schematic overview of the workflow search space in our framework. The search space consists of various sequential sets of algorithms, where each algorithm may include various hyperparameters, as indicated by the leaves in the trees. An example of a workflow, i.e., a specific combination of algorithms and parameters, is indicated by the gray nodes. Abbreviations: AdaBoost: adaptive boosting; ADASYN: adaptive synthetic sampling; KNN: k-nearest neighbor; GLCM: gray level co-occurence matrix; SMOTE: synthetic minority oversampling technique; SVM: support vector machine.}
		\label{fig: schematic}
	\end{figure*}

	A schematic overview of the algorithm and hyperparameter search space is shown in \autoref{fig: schematic}. Including new algorithms and hyperparameters in this reformulation is straight-forward, as these can simply be added to $\mathcal{A_C}$ and $\boldsymbol{\Lambda}_C$, respectively.
	
	As a loss function $\mathcal{L}$, we use the weighted  $F_{1,w}$ score, which is the harmonic mean of precision and recall, and thus a class-balanced performance metric:
	
	\begin{equation}
		F_{1, w} = 2 \sum_{c=1}^{n_{\text{classes}}} \frac{N_c}{N_{\text{total}}} \frac{\text{PREC}_c \times \text{REC}_c}{\text{PREC}_c + \text{REC}_c},
		\label{eq: F1}
	\end{equation}
	where the number of classes $n_{\text{classes}} = 2$ for binary classification, $N_c$ the number of samples of class $c$, $N_{\text{total}}$ the total number of samples, and $\text{PREC}_c$ and $\text{REC}_c$ the precision and recall of class $c$, respectively. For the binary classification problems in this study, the $F_{1,w}$ for the positive label is used.
	
	\subsubsection{CASH optimization strategies}
	We compare two strategies for solving the CASH optimization problem. 
	
	First, we use a random search algorithm, as it is efficient and often performs well, making it a popular baseline \cite{RN84}. In random search, $N_{\text{RS}}$ workflows are randomly sampled from the search space $\boldsymbol{\Lambda}_C$, and their $F_{1,w}$ scores are calculated.
	
	Second, we use Bayesian optimization, in particular Sequential Model-based Algorithm Configuration ({\toolbox $ SMAC $})  \cite{lindauer2022smac3}, as it was originally used to solve the CASH optimization problem \cite{RN56}, and in AutoML is the state-of-the-art and one of the most popular methods \cite{RN694, RN1358}, e.g., used in many of the winning workflows in the ChaLearn challenge \cite{RN1245} and in prominent AutoML systems such as {\toolbox{Auto-WEKA}} \cite{RN55} and {\toolbox{Auto-sklearn}} \cite{RN70}. Moreover, Bayesian optimization is suitable to optimize the {\toolbox{WORC}} search space as it facilitates both categorical and continuous optimization, large search spaces, and combined algorithm selection and hyperparameter selection. SMAC uses a random search as initialization strategy, a random forest as surrogate model to relate the blackbox hyperparameters to the output, and the expected improvement as acquisition function to determine which blackbox hyperparameter configurations should be tried in the next iteration. The optimization target and evaluation strategy are the same as used by the random search, i.e., optimizing the mean $F_{1, w}$ on the validation datasets in a cross-validation with $k_{\text{training}}$ iterations. 
	
	
	\subsubsection{Ensembling}
	In radiomics studies comparing multiple workflows there is often not a clear winner in terms of overall performance. However, the prediction for individual samples may considerably vary per workflow. Moreover, due to the CASH optimization, we expect the best performing workflow likely to overfit. Hence, we hypothesize that by combining different workflows in an ensemble, the performance and generalizability of radiomics models will be improved \cite{RN70}. Additionally, ensembling may serve as a form of regularization, as local minima in the CASH optimization are balanced by other workflows in the ensemble. We therefore hypothesize that ensembling will also lead to a more stable solution of the CASH optimization. 
	
	Therefore, instead of selecting the single best workflow, we propose to use an ensemble $\mathcal{E}$, for which we evaluate three strategies.
	
	First, as optimizing the ensemble construction on the training dataset may in itself lead to overfitting, we propose a simple approach which does not require any fitting: combining a fixed number $N_{\text{ens}}$ of the best performing workflows by averaging their predictions (i.e., the posterior probabilities for binary classification). The workflows are ranked based on their mean validation $F_{1,w}$.
	
	Second, instead of using a fixed number of workflows, we optimize this number by selecting $N_{\text{ens}} \in \{1, \ldots, 100\}$ with the highest mean validation $F_{1, w}$. This method will be referred to as ``FitNumber''. 
	
	Lastly, we use the often-cited approach from \cite{RN69}, which is also the default method for {\toolbox{Auto-sklearn}} \cite{RN70}. The approach starts with an empty ensemble and iteratively performs forward selection to select which estimator gives the largest improvement, with repetition of the same estimator to allow a form of weighting. To prevent overfitting, bagging is applied, where in each bagging iteration only a subset of the available models is included. This method will be referred to as ``ForwardSelection''. 
	
	To limit the computational burden, the maximum number of workflows for the FitNumber and ForwardSelection methods was set at $N_{\text{ens}} = 100$. 
	
	\subsubsection{The {\toolbox{WORC}} optimization algorithm}
	The {\toolbox{WORC}} optimization algorithm is depicted in Algorithm \autoref{alg: WORC}: for simplicity, the version with random search as optimization strategy and $N_{\text{ens}}$ ensembling is depicted. All optimization is performed on the training dataset by using a stratified random-split cross-validation with $k_{\text{training}}=5$, using 80\% for training and 20\% for validation in a stratified manner. Random-split cross-validation is used as this allows a fixed ratio between the training and validation datasets independent of $k_{\text{training}}$, and is consistent with our evaluation setup (\autoref{sec: MethodsEvaluation}). The algorithm returns a trained ensemble $\mathcal{E}$.
	
	\begin{algorithm}
		\caption{The {\toolbox{WORC}} optimization algorithm when using random search as optimization strategy and $N_{\text{ens}}$ ensembling.}
		\begin{algorithmic}[1]
			\REQUIRE {$\boldsymbol{\Lambda}_C, N_{\text{RS}}, k_{\text{training}}, N_{\text{ens}}$}
			\FOR{$n \in \{1, \ldots, N_{\text{RS}} \}$}
			\STATE $\boldsymbol{\lambda}_n \gets \text{Random} \left(\boldsymbol{\Lambda}_C \right)$
			\STATE $\mathcal{L}_n = \frac{1}{k_{\text{training}}} \sum_{i=1}^{k_{\text{training}}} \mathcal{L} \left(\boldsymbol{\lambda_n}, \mathcal{D}^{(i)}_{\text{train}}, \mathcal{D}^{(i)}_{\text{valid}} \right)$
			\ENDFOR
			\STATE $\boldsymbol{\Lambda}_{ranked} \gets \text{Rank}(\{\boldsymbol{\lambda}_1, \ldots, \boldsymbol{\lambda}_{N_{\text{RS}}}\} \propto \{\mathcal{L}_1, \ldots, \mathcal{L}_{N_{\text{RS}}} \})$
			\STATE $\boldsymbol{\Lambda}_{\text{ens}} \gets \boldsymbol{\Lambda}_{ranked} \left[1 : N_{\text{ens}} \right] $
			\STATE Retrain $\boldsymbol{\Lambda}_{\text{ens}}$ on full training set
			\STATE Combine $\boldsymbol{\Lambda}_{\text{ens}}$ into ensemble $\mathcal{E}$
			\STATE \textbf{return} $\mathcal{E}$
		\end{algorithmic}
		\label{alg: WORC} 
	\end{algorithm}
	
	The {\toolbox{WORC}} toolbox is implemented in Python3 and available open-source \cite{RN63}\footnote{This DOI represents all versions, and will always resolve to the latest one.} under the Apache License, Version 2.0. Documentation on the {\toolbox{WORC}} toolbox can be found online \cite{RN540}, and several tutorials are available\footnote{\href{https://github.com/MStarmans91/WORCTutorial}{https://github.com/MStarmans91/WORCTutorial}}. A minimal working example is shown in Algorithm \autoref{alg: WORCCodeExample}.
	
	\subsection{Radiomics workflow search space} \label{sec: Methods_Components}
	In order to formulate radiomics as CASH optimization problem, the workflow needs to be modular and consist of standardized components. In this way, for each component, a set of algorithms and hyperparameters can be included. We therefore split the radiomics workflow into the following components: image and segmentation preprocessing, feature extraction, feature and sample preprocessing (consisting of multiple components itself, e.g., feature selection, resampling), and machine learning. An overview of the default included components, algorithms, and associated hyperparameters in the {\toolbox{WORC}} framework is provided in \autoref{table: hyperparameters}; more details are given in \refappendix{sec: SupplementaryMaterial1}.
	
	\begin{table}
		\centering
		\scalebox{0.925}{ 
			\begin{tabular}{l l l}
				\toprule
				\textbf{Algorithm} & \textbf{Hyperparameter} & \textbf{Distribution} \\
				\hline
				\emph{1: Feature Selection} & & \\
				
				Group-wise selection & \textit{Activator} & $\mathcal{B}(1.0)$ \\
				& \textit{Activator} per group & $17 \times \mathcal{B}(0.5)$ \\
				\hline
				\emph{2: Feature Imputation} & \textit{Selector} & $\mathcal{C}(5)$ \\
				
				Mean & - & - \\
				Median & - & - \\
				Mode & - & - \\
				Constant (zero) & - & - \\
				k-nearest neighbors & Nr. Neighbors & $\mathcal{U}^d(5, 10)$ \\
				\hline
				\emph{3: Feature Selection} &&\\
				
				Variance Threshold & \textit{Activator} & $\mathcal{B}(1.0)$ \\
				\hline
				\emph{4: Feature Scaling} &&\\
				
				Robust z-scoring & - & - \\
				\hline
				\emph{5: Feature Selection} &&\\
				
				RELIEF & \textit{Activator} & $\mathcal{B}(0.2)$ \\
				& Nr. Neighbors & $\mathcal{U}^d(2, 6)$ \\
				& Sample size & $\mathcal{U}(0.75, 0.95)$ \\
				& Distance P & $\mathcal{U}^d(1, 4)$ \\
				& Nr. Features & $\mathcal{U}^d(10, 50)$ \\
				\hline
				\emph{6: Feature Selection} &&\\
				
				SelectFromModel & \textit{Activator} & $\mathcal{B}(0.2)$ \\
				& Type & $\mathcal{C}(3)$ \\
				& LASSO alpha & $\mathcal{U}(0.1, 1.5)$ \\
				& RF Nr. Trees & $\mathcal{U}^d(10, 100)$ \\
				\hline
				\emph{7: Dimensionality Reduction} &&\\
				
				Principal component analysis & \textit{Activator} & $\mathcal{B}(0.2)$ \\
				& Type & $\mathcal{C}(4)$ \\
				& & \\
				\hline
				\emph{8: Feature Selection} &&\\
				
				Univariate testing & \textit{Activator} & $\mathcal{B}(0.2)$ \\
				& Threshold & $\mathcal{U}^l(10^{-3}, 10^{-2.5})$ \\
				\hline
				\emph{9: Resampling} & \textit{Activator} & $\mathcal{B}(0.2)$ \\
				
				& \textit{Selector} & $\mathcal{U}^d(1, 6)$ \\
				RandomUnderSampling & Strategy & $\mathcal{C}(4)$ \\
				RandomOverSampling & Strategy & $\mathcal{C}(4)$ \\
				NearMiss & Strategy & $\mathcal{C}(4)$ \\
				NeighborhoodCleaningRule & Strategy & $\mathcal{C}(4)$ \\
				& Nr. Neighbors & $\mathcal{U}^d(3, 15)$ \\
				& Cleaning threshold & $\mathcal{U}(0.25, 75)$ \\
				SMOTE & Type & $\mathcal{C}(4)$ \\
				& Strategy & $\mathcal{C}(4)$ \\
				& Nr. Neighbors & $\mathcal{U}^d(3, 15)$ \\
				ADASYN & Strategy & $\mathcal{C}(4)$ \\
				& Nr. Neighbors & $\mathcal{U}^d(3, 15)$ \\
				\hline
				\emph{10: Classification} & \textit{Selector} & $\mathcal{U}^d(1, 8)$\\
				
				Support vector machine & Kernel & $\mathcal{C}(3)$ \\
				& Regularization & $\mathcal{U}^l(10^0, 10^6)$ \\
				& Polynomial degree & $\mathcal{U}^d(1, 7)$ \\
				& Homogeneity & $\mathcal{U}(0, 1)$ \\
				& RBF $\gamma$ & $\mathcal{U}^l(10^{-5}, 10^{5})$ \\
				Random forest & Nr. Trees & $\mathcal{U}^d(10, 100)$ \\
				& Min. samples / split & $\mathcal{U}^d(2, 5)$ \\
				& Max. depth & $\mathcal{U}^d(5, 10)$ \\
				Logistic regression & Regularization & $\mathcal{U}(0.01, 1)$ \\
				& Solver & $\mathcal{C}(2)$ \\
				& Penalty & $\mathcal{C}(3)$ \\
				& $L_1$-ratio & $\mathcal{U}(0, 1)$ \\
				Linear discriminant analysis & Solver & $\mathcal{C}(3)$ \\
				& Shrinkage & $\mathcal{U}^l(^10^{-5}, 10^5)$ \\
				Quadratic discriminant analysis & Regularization & $\mathcal{U}^l(^10^{-5}, 10^5)$ \\
				Gaussian Naive Bayes & Regularization & $\mathcal{U}(0, 1)$ \\
				AdaBoost & Nr. Estimators & $\mathcal{U}^d(10, 100)$ \\
				& Learning rate & $\mathcal{U}^l(0.01, 1)$ \\
				XGBoost & Nr. Rounds & $\mathcal{U}^d(10, 100)$ \\
				& Max. depth & $\mathcal{U}^d(3, 15)$ \\
				& Learning rate & $\mathcal{U}^l(0.01, 1)$ \\
				& $\gamma$ & $\mathcal{U}(0.01, 10)$ \\
				& Min. child weight & $\mathcal{U}^d(1, 7)$ \\
				& \% Random samples & $\mathcal{U}(0.3, 1.0)$ \\
				\bottomrule\\
		\end{tabular}}
		\caption{Overview of the algorithms and associated hyperparameter search space as used in the {\toolbox{WORC}} framework for binary classification problems. Definitions: $\mathcal{B}(p)$: Bernoulli distribution, equaling value $True$ with probability $p$; $\mathcal{C}(c)$ a categorical distribution over $c$ categories; $\mathcal{U}(\text{min}, \text{max})$: uniform distribution; $\mathcal{U}^d(\text{min}, \text{max})$: uniform distribution with only discrete values; $\mathcal{U}^l(\text{min}, \text{max})$: uniform distribution on a logarithmic scale. Abbreviations: RBF: radial basis function.}
		\label{table: hyperparameters}
	\end{table}
	
	An additional problem is that there is no clearly defined range of radiomics solutions to choose from, i.e., search space. Most radiomics reviews are not systematic but rather some examples \cite{Li2022, Bera2022, RN261, RN691, RN436, RN674, RN798, RN799, RN35, RN800, RN1082, RN1079, RN1227, RN447, RN505}. \cite{RN686} perform a systematic review, but only present some of the most commonly used machine learning methods, not all radiomics components, and do not include how algorithms were exactly used, e.g., stand-alone or in combination with other algorithms, as feature selection or classification step when they can fulfill both tasks, and their hyperparameter configuration or optimization.
	
	Hence, to address the issue of manual optimization, we also need to design an adequate search space. To this end, for each component, we have included a large collection of the algorithms and associated hyperparameters most commonly included in the radiomics literature. Hyperparameter search spaces were either set based on recommendations by the used software or the literature, set to span a wide range of logical values (e.g., $[0, 1] $ for ratios), or otherwise uniformly distributed around a default (logarithmic) value of the algorithm.
	
	\subsection{Fingerprinting}\label{sec: Fingerprinting}
	While it is generally difficult to \textit{a priori} determine which radiomics algorithms or hyperparameters may work best or not at all on a specific dataset, there may be some obvious relations. To exploit such prior knowledge, we introduce a light fingerprinting mechanism inspired by \cite{RN714} to reduce the search space. 
	
	First, we distinguish between qualitative modalities, i.e., images that do not have a fixed unit and scale (e.g., ultrasound, qualitative MRI such as T1-weighted) and quantitative modalities (e.g., CT, quantitative MRI such as T1 mapping). Only in qualitative images, image normalization is applied. Furthermore, when discretization is required for feature computation (e.g., gray level matrix features), a fixed bin count is used in qualitative modalities, while a fixed bin width is used in quantitative modalities \cite{RN539, RN761}.
	
	Second, we inspect the mean pixel spacing and slice thickness in a dataset to choose between using 2D, 2.5D, and/or 3D feature extraction. Many radiomics studies include datasets with variations in the slice thickness due to heterogeneity in the acquisition protocols. This may cause feature values to be dependent on the acquisition protocol. Moreover, the slice thickness is often substantially larger than the pixel spacing, resulting in a ``2.5D'' representation rather than 3D. We therefore define three scenarios: 1) If the images and/or segmentations in a dataset consist of a single 2D slice, only 2D features are used; 2) If the mean of all images is (almost) isotropic, i.e., slice thickness is similar to pixel spacing, 3D features are extracted. In this case, a 2.5D approach is used for features that are only defined in 2D; and 3) If the mean of all images is anisotropic, features are extracted in 2.5D. To extract 2.5D features, the features are extracted per 2D slice and aggregated over all slices, after which various first order statistics are computed. We use a conservative definition for similarity between slice thickness and pixel spacing:  $\text{slice thickness} \leq 2 \times \text{pixel spacing}$.
	
	Third, if the dataset is relatively balanced, resampling methods are deemed not useful and are omitted. We define relatively balanced conservatively as maximally a $60/40$ ratio between the classes. 
	
	\subsection{Statistics} \label{sec: MethodsEvaluation}
	Evaluation using a single dataset is performed through random-split cross-validation with $k_{\text{test}}=100$, see \autoref{fig: crossval}(a). Random-split cross-validation, also known as Monte Carlo cross-validation, was chosen as it has a relatively low computational complexity while facilitating generalization error estimation \cite{RN866, RN585}. In each iteration, the data is randomly split in 80\% for training and 20\% for testing in a stratified manner. In each random-split iteration, all CASH optimization is performed within the training set according to Algorithm \autoref{alg: WORC} to eliminate any risk of overfitting on the test set. When a fixed, independent training and test set are used, only the internal random-split cross-validation with $k_{\text{training}}=5$ on the training set for the CASH optimization is used, see \autoref{fig: crossval}(b).
	
	Performance metrics used for evaluation of the test set include the Area Under the Curve (AUC), calculated using the Receiver Operating Characteristic (ROC) curve, Balanced Classification Rate (BCR) \cite{RN704}, $F_{1, w}$, sensitivity, and specificity. When a single dataset is used, and thus $k_{\text{test}}=100$ random-split cross-validation, metric 95\% confidence intervals are constructed using the corrected resampled t-test, thereby taking into account that samples in the cross-validation splits are not statistically independent \cite{RN585}. When a fixed training and test set are used, 95\% confidence intervals are constructed using 1000x bootstrap resampling of the test dataset and the standard method for normal distributions (\cite{RN697}, table 6, method 1). ROC confidence bands are constructed using fixed-width bands \cite{RN117}.
	
	\section{Experiments} \label{sec: datasets}
	\subsection{Datasets from twelve different clinical applications}
	In order to evaluate our {\toolbox{WORC}} framework, experiments were performed on twelve different clinical applications: see \autoref{tab: datasets} for an overview including example images. We focused on oncology applications as these are most common in radiomics, but also included one widely studied non-oncology application (Alzheimer). For the oncology applications, we used routinely collected, clinically representative, multi-center datasets to train and evaluate our method. This facilitates generalization of the resulting biomarkers across image acquisition protocols and thus across clinical centres, increasing the feasibility of applying such a biomarker in routine clinical practice.
	
	\begin{table*}
		\centering
		
		\begin{minipage}{\textwidth}
			\centering
			\includegraphics[width=0.975\textwidth]{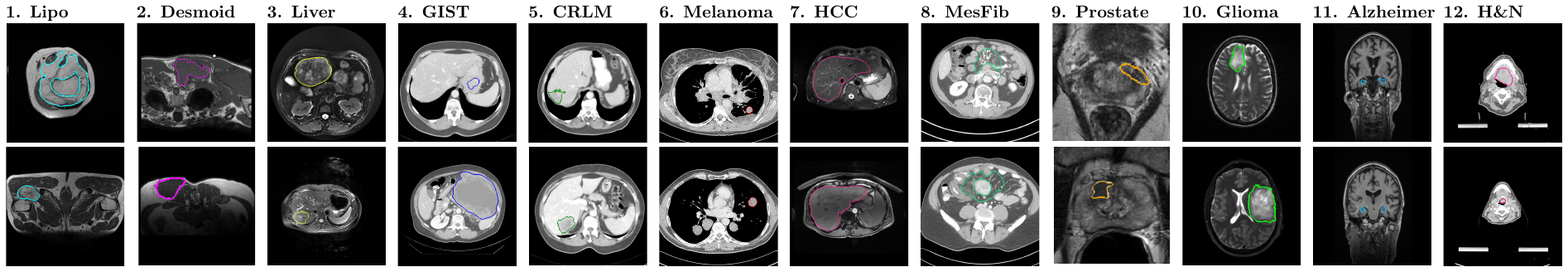}
			\vspace{0.25cm} 
		\end{minipage}
		
		\begin{tabular}{l l | l p{1.5cm} p{2cm} p{9cm}}
			\toprule
			\textbf{\#} & \textbf{Dataset} & \textbf{Patients} & \textbf{Modality} & \textbf{Segmentation} & \textbf{Description} \\
			\midrule
			1. & Lipo$^O$ & 115 & T1w MRI & Tumor & Distinguishing well-differentiated liposarcoma from lipoma in 116 lesions from 115 patients \cite{RN574}. \\
			\hdashline
			2. & Desmoid$^O$ & 203 & T1w MRI & Tumor & Differentiating desmoid-type fibromatosis from soft-tissue sarcoma \cite{RN678}. \\
			\hdashline
			3. & Liver$^O$ & 186 & T2w MRI & Tumor & Distinguishing malignant from benign primary solid liver lesions \cite{starmans2024BLT}. \\
			\hdashline
			4. & GIST$^O$ & 246 & CT & Tumor & Differentiating gastrointestinal stromal tumors (GIST) from other intra-abdominal tumors in 247 lesions from 246 patients \cite{RN1246}. \\
			\hdashline
			5. & CRLM$^O$ & 77 & CT & Tumor & Distinguishing replacement from desmoplastic histopathological growth patterns in colorectal liver metastases (CRLM) in 93 lesions from 77 patients \cite{CRLMPaper}. \\
			\hdashline
			6. & Melanoma$^O$ & 103 & CT & Tumor & Predicting the \textit{BRAF} mutation status in melanoma lung metastases in 169 lesions from 103 patients \cite{RN791}. \\
			\hdashline
			7. & HCC & 154 & T2w MRI & Liver & Distinguishing livers in which no hepatocellular carcinoma (HCC) developed from livers with HCC at first detection during screening \cite{RN706}. \\
			\hdashline
			8. & MesFib & 68 & CT & Surrounding mesentery & Identifying patients with mesenteric fibrosis at risk of developing intestinal complications \cite{RN923}.\\
			\hdashline
			9. & Prostate & 40 & T2w MRI, DWI, ADC & Lesion & Classifying suspected prostate cancer lesions in high-grade (Gleason $>6$ ) versus low-grade (Gleason $<=6$) in 72 lesions from 40 patients \cite{RN480}. \\
			\hdashline
			10. & Glioma & 413 & T1w \& T2w MRI & Tumor & Predicting the 1p/19q co-deletion in patients with presumed low-grade glioma with a training set of 284 patients and an external validation set of 129 patients \cite{RN492}. \\
			\hdashline
			11. & Alzheimer & 848 & T1w MRI & Hippocampus & Distinguishing patients with Alzheimer's disease from cognitively normal individuals in 848 subjects based on baseline T1w MRIs \cite{RN1103}. \\
			\hdashline
			12. & H\&N & 137 & CT & Gross tumor volume & Predicting the T-stage (high ($\geq 3$) or low ($<3$)) in patients with head-and-neck cancer \cite{RN681}. \\
			\bottomrule
			\multicolumn{6}{l}{$^O$Dataset publicly released as part of this study \cite{WORCDatabase}.}
		\end{tabular}
		\caption{Overview of the twelve datasets used in this study to evaluate our {\toolbox{WORC}} framework, including examples of the 2D slices from the 3D imaging data. For each dataset, for one patient of each of the two classes, the 2D slice in the primary scan direction (e.g., axial) with the largest area of the segmentation is depicted; the boundary of the segmentation is projected in color on the image.Abbreviations: ADC: Apparent Diffusion Coefficient; CT: Computed Tomography; DWI: Diffusion Weighted Imaging; MRI: Magnetic Resonance Imaging; T1w: T1 weighted; T2w: T2 weighted.}
		\label{tab: datasets}
	\end{table*}
	
	The first six datasets (Lipo, Desmoid, Liver, GIST, CRLM, and Melanoma) are publicly released as part of this study, see \cite{WORCDatabase}. Three private datasets (HCC, MesFib, and Prostate) could not be made publicly available. The final three datasets (Glioma, Alzheimer, and H\&N) were already publicly available \cite{RN492, RN1104, RN681}.
	
	For the Glioma dataset, the raw imaging data was not available. Instead, pre-computed radiomics features, age, and sex were available \cite{RN707}, which were directly used.
	
	The Alzheimer dataset was obtained from the Alzheimer’s Disease Neuroimaging Initiative (ADNI) database\footnote{The ADNI was launched in 2003 as a public-private partnership, led by Principal Investigator Michael W. Weiner, MD. The primary goal of ADNI was to test whether serial MRI, positron emission tomography (PET), other biological markers, and clinical and neuropsychological assessment can be combined to measure the progression of mild cognitive impairment (MCI) and early Alzheimer’s disease (AD). For up-to-date information, see \href{www.adni-info.org}{www.adni-info.org}.}. Here, radiomics was used to distinguish patients with AD from cognitively normal (CN) individuals. The cohort as described by \cite{RN1103} was used, which includes 334 patients with AD and 520 CN individuals, with approximately the same mean age in both groups (AD: 74.9 years, CD: 74.2 years). The hippocampus was chosen as region of interest as it is known to suffer from atrophy early in the disease process of AD, which was automatically segmented \cite{RN708}.
	
	The H\&N dataset \cite{RN681} was obtained from a public database\footnote{\href{https://xnat.health-ri.nl/data/projects/stwstrategyhn1}{https://xnat.health-ri.nl/data/projects/stwstrategyhn1}}. For each lesion, the first gross tumor volume (GTV-1) segmentation was used as region of interest. Patients without a CT scan or a GTV-1 segmentation were excluded.
	
	For each experiment, per patient, one or more scan(s) and segmentation(s) or a feature set (Glioma), and a ground truth label are provided. All scans were made at ``baseline'', i.e., before any form of treatment or surgery. The Glioma dataset consists of fixed, independent training and test sets and is thus evaluated using 1000x bootstrap resampling. In the other eleven datasets, the performance is evaluated using the default $k_{\text{test}}=100$ random-split cross-validation.
	
	\subsection{Workflow optimization: comparison of random search, Bayesian optimization, and ensembling settings}\label{sec: MethodsRSEns}
	First, several experiments were conducted to compare the performance of random search and Bayesian optimization with various ensembling approaches. For reproducibility, these experiments were performed using the six datasets publicly released in this study.
	
	To investigate the influence of the computational budget, we varied $N_{\text{RS}} \in \{10, 50, 100, 1000, 10000, 25000\}$. As random search can become unstable at few iterations, we repeated each experiment ten times with different seeds for the random number generator to evaluate its stability. Random search is highly efficient \cite{RN84}, while {\toolbox SMAC} requires time for the surrogate model fitting and can thus explore less workflows in the same computation time. We hypothesize that the random search will outperform {\toolbox SMAC} on a low computational budget, as the surrogate model will not be highly accurate and relatively few function evaluations can be performed, but that {\toolbox SMAC} will outperform the random search at higher computational budget as these disadvantages decrease. We therefore compare the random search with three different computational budgets for {\toolbox SMAC}: low (36 hours), medium (355 hours) and high (1842 hours), roughly corresponding with the computational budgets for $N_{\text{RS}} \in \{1000, 10000, 50000\}$, respectively. These relatively high budgets compared to random search were chosen to accommodate for the surrogate model fitting. The {\toolbox SMAC} runtime is defined as the wallclock time spent on fitting and evaluating the workflows, fitting the surrogate model, and selecting new configurations to execute. Note that this is the computational budget for the optimization in a single train-test cross-validation iteration. All SMAC experiments were also repeated ten times with different random number generator seeds.
	
	Both random search and {\toolbox SMAC} optimize $F_{1, w}$ on the validation sets. An improved validation performance may not necessarily translate to better generalization, i.e., an increased test performance. Thus, to evaluate which method performs better in terms of both generalization and optimization, we compare not only the test set performance between random search and {\toolbox SMAC}, but also the optimization criterion, i.e., the $F_{1, w}$ of the single best found workflow on the validation set.
	
	The three different ensembling methods, $N_{\text{ens}}$, FitNumber, and ForwardSelection, were compared for each $N_{\text{RS}}$ and {\toolbox SMAC} runtime setting. For the $N_{\text{ens}}$ method, we hypothesize that increasing $N_{\text{ens}}$ at first will improve the performance and stability, and after some point, when $N_{\text{ens}}$/$N_{\text{RS}}$ becomes too high, will reduce the performance and stability as bad workflows are added to the ensemble. We therefore varied the ensemble size $N_{\text{ens}} \in \{1 \text{(i.e., no ensembling)}, 10, 50, 100 \}$.
	
	For each configuration, both the average performance (mean $F_{1,w}$) and the stability (standard deviation of $F_{1,w}$) over the ten repetitions were assessed on the test set. To limit the computational burden, $k_{\text{test}}=20$ was used instead of the default $k_{\text{test}}=100$, and the $N_{RS}=25000$ experiment was only performed once instead of ten times.
	
	Details on the used hardware and computation time are described in \refappendix{sec: SupplementaryMaterial2Computationtime}.
	
	\subsection{Comparison to radiomics baseline} 
	We additionally compared the {\toolbox{WORC}} algorithm to a radiomics baseline on the six datasets publicly released in this study. As mentioned in \autoref{sec: Methods_Components}, a variety of methods is used in radiomics, hence ``the state-of-the-art'' in radiomics is not well-defined. According to \cite{RN686}, the most commonly used choices for the individual radiomics components are {\toolbox PyRadiomics} \cite{RN761} for feature extraction, LASSO \cite{RN514} for feature selection, and logistic regression for classification. Various examples of combinations of these components can also be found in the literature \cite{RN1239, RN1234, RN1235, RN1236, RN1237, RN1238}.
	
	Hence, to compare the {\toolbox{WORC}} algorithm to a radiomics baseline, additional experiments were conducted using these selected methods ({\toolbox PyRadiomics} + LASSO + logistic regression). As both LASSO and the logistic regression have hyperparameters that require tuning, we used the same random search as the {\toolbox{WORC}} algorithm to optimize their settings. As the search space is small, $N_{RS}=1000$ was used. Effectively, this corresponds with substantially limiting the {\toolbox{WORC}} search space, thus resulting in a similar computation time. While the {\toolbox{WORC}} algorithm has to both perform method selection and hyperparameter optimization, the radiomics baseline can dedicate all computational budget to hyperparameter optimization and thus focus on finetuning the selected method configurations. The same ensembling approaches were compared as in above experiments.
	
	\subsection{Validation of optimization algorithm default settings}
	Based on above experiments, the default optimization and ensemble strategies for the {\toolbox{WORC}} optimization algorithm were determined and evaluated on all twelve datasets using the default $k_{\text{test}}=100$.
	
	\section{Results} \label{sec: res}
	\subsection{Comparing random search, Bayesian optimization, ensembling settings, and a radiomics baseline}
\begin{figure*}
		\centering
		\includegraphics[width=0.975\textwidth]{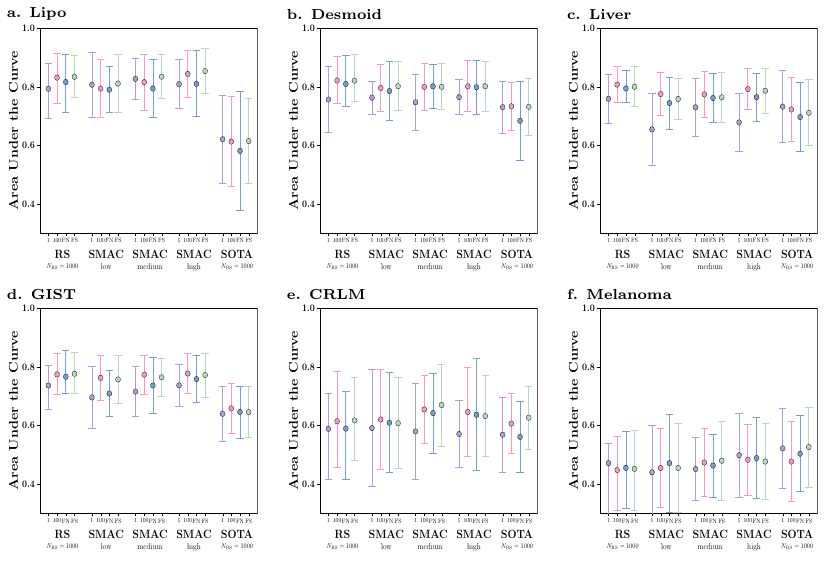}
		\caption{Error plots of the area under the receiver operating characteristic curve (AUC) on the test datasets of the radiomics models on six datasets (Lipo, Desmoid, Liver, GIST, CRLM, and Melanoma) for two optimization strategies (RS: random search with $N_{RS}=1000$, SMAC: sequential model-based algorithm configuration with different computational budgets (low, medium, high)) and a radiomics state-of-the-art (SOTA) baseline, when using either the single best found validation workflow (1) or one of three ensembling strategies (100: $N_{\text{ens}} = 100$, FN: FitNumber, FS: ForwardSelection). The error plots represent 95\% confidence intervals, estimated through $k_{\text{test}}=20$ random-split cross-validation on the entire dataset. The circle represents the mean.}
		\label{fig: errorplot_AUC_test}
	\end{figure*}

	The test AUC of the {\toolbox WORC} algorithm for various random search, Bayesian optimization, and ensembling settings, and the radiomics baseline in the six public datasets, is depicted in \autoref{fig: errorplot_AUC_test}. For brevity, only $N_{RS}=1000$ for the random search, and not all ensemble settings are shown: detailed results for varying $N_{\text{RS}}$ and $N_{\text{ens}}$ for the random search are reported in \autoref{tab: rsens}.
	
	\subsubsection{Random search and ensembling}
	As seen in \autoref{tab: rsens}, for five out of six datasets in this experiment (Lipo, Desmoid, Liver, GIST, and CRLM), the mean test performance generally improved when increasing both $N_{\text{RS}}$ and $N_{\text{ens}}$ in the $N_{\text{ens}}$ method. For the sixth dataset (Melanoma), the performance for varying $N_{\text{RS}}$ and $N_{\text{ens}}$ was similar. This exception can be attributed to the fact that it is the only dataset in this study where we could not successfully construct a predictive model. 
	
	More specifically, in the first five datasets, the mean test $F_{1, w}$ for the highest values, $N_{\text{RS}}=25000$ and $N_{\text{ens}}=100$, was substantially higher for all five datasets than at the lowest values, $N_{\text{RS}}=1$ (i.e., only trying one random workflow) and $N_{\text{ens}}=1$ (i.e., no ensembling), (Lipo: 0.75 vs. 0.69; Desmoid: 0.76 vs. 0.70; Liver: 0.71 vs. 0.64; GIST: 0.71 vs. 0.63; and CRLM: 0.57 vs. 0.53). The mean test $F_{1, w}$ of $N_{\text{RS}}=1000$ was very similar to that of $N_{\text{RS}}=25000$, while the latter takes 25 times longer to execute. This indicates that at some point, here $N_{\text{RS}}=1000$, increasing the computation time by trying out more workflows does not result in a test performance increase anymore.
	
	At $N_{\text{RS}}=10$ and $N_{\text{ens}}=1$, the standard deviation of the test $F_{1, w}$ (Lipo: 0.024; Desmoid: 0.018; Liver: 0.017; GIST: 0.010; and CRLM: 0.022) was substantially higher than at $N_{\text{RS}}=10000$, $N_{\text{ens}}=100$ (Lipo: 0.005; Desmoid: 0.003; Liver: 0.004; GIST: 0.003; and CRLM: 0.009). This indicates that increasing $N_{\text{RS}}$ and $N_{\text{ens}}$ improves the stability of the model. The standard deviations at $N_{\text{RS}}=10000$ were similar to those at $N_{\text{RS}}=1000$, illustrating that, similar to the mean performance, the stability at some point converges. The confidence intervals for the different ensemble methods showed substantial overlap, but on average the $N_{\text{ens}} = 100$ ensembling method consistently outperformed the FitNumber and ForwardSelection methods on all datasets for $N_{\text{RS}} \leq 1000$, and showed similar performance on lower settings.
	
	\subsubsection{Bayesian optimization and ensembling}
	As seen in \autoref{fig: errorplot_AUC_test}, compared with $N_{\text{RS}} = 1000$ random search, {\toolbox SMAC} generally performed slightly worse in terms of test AUC when not using ensembling. When using ensembling, {\toolbox SMAC} still performed slightly worse or similar on all computational budgets to that of the random search, as the confidence intervals showed substantial overlap. Hence, while requiring substantially more computational budget on the ``medium'' and ``high'' budgets, Bayesian optimization through {\toolbox SMAC} did not yield performance improvements.
	
	The validation $F_{1, w}$ for all methods, which the random search, {\toolbox SMAC}, and the ensembling optimize, is shown in \autoref{fig: errorplot_F1_validation}. For the single best found workflow, while on the ``low'' budget {\toolbox SMAC} performed slightly worse than random search, it performed substantially better on the higher budgets. Combined with the ForwardSelection or FitNumber ensembling methods, both {\toolbox SMAC} and random search reached near perfect validation $F_{1, w}$. Compared with the above test AUC, this clearly shows both {\toolbox SMAC} and the two advanced ensembling methods are overfitting on the validation dataset without resulting in an increased performance on the test dataset. Contrarily, while $N_{\text{ens}} = 100$ ensembling, which does not apply any fitting, resulted in a lower validation $F_{1, w}$, the test AUC was substantially higher than the single best found workflow for both {\toolbox SMAC} and the random search.
	
	
	\subsubsection{Radiomics baseline}
	As seen in \autoref{fig: errorplot_AUC_test}, in terms of mean test AUC, the radiomics baseline ({\toolbox PyRadiomics} + LASSO + logistic regression) performed worse than the {\toolbox{WORC}} algorithm both when using random search and {\toolbox SMAC} in five datasets (Lipo: 0.62, Desmoid: 0.73, Liver: 0.72, GIST: 0.64, CRLM: 0.57). On the Melanoma (0.52) dataset, where no method performed well, the performance was slightly better. Additionally, while both random search and {\toolbox SMAC} substantially benefited from ensembling, the radiomics baseline performance remained similar. Hence, the {\toolbox{WORC}} algorithm substantially outperformed the radiomics baseline.
	
	\subsection{Application of final default {\toolbox{WORC}} algorithm to twelve datasets}
	Based on the above results, the final default {\toolbox{WORC}} algorithm uses a random search with $N_{\text{RS}}=1000$ and $N_{\text{ens}} = 100$ ensembling. This approach performed well while having a relatively low computational budget.
	
\begin{figure}
		\centering
		\includegraphics[width=0.475\textwidth]{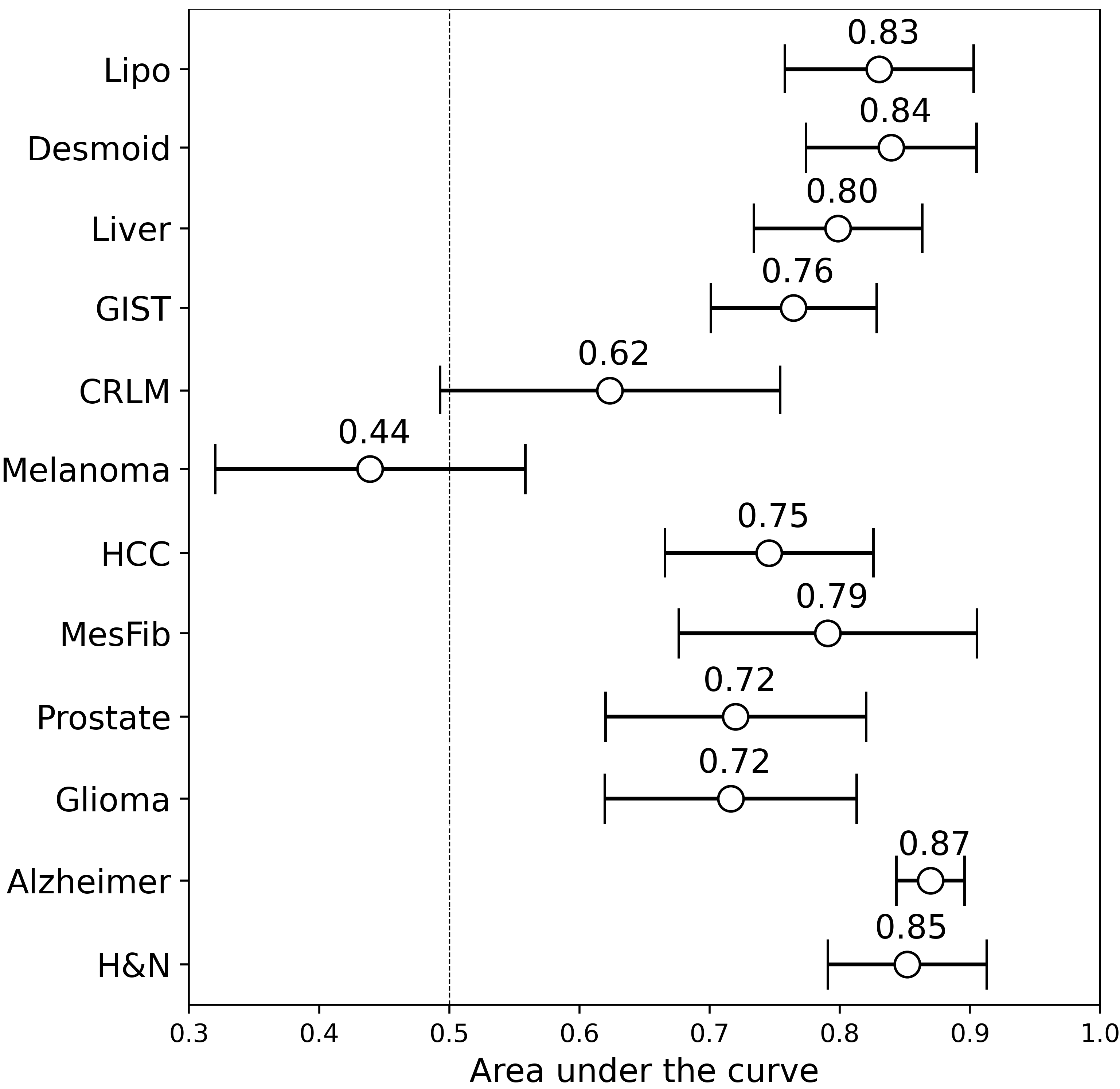}
		\caption{Error plots of the area under the receiver operating characteristic curve (AUC) of the radiomics models on twelve datasets. The error plots represent the 95\% confidence intervals, estimated through $k_{\text{test}}=100$ random-split cross-validation on the entire dataset (all except Glioma) or through 1000x bootstrap resampling of the independent test set (Glioma). The circle represents the mean (all except Glioma) or point estimate (Glioma), which is also stated to the right of each circle. The dashed line corresponds to the AUC of random guessing (0.50).}
		\label{fig: AUCBoxplot}
	\end{figure}

	Error plots of the AUC from the application of our {\toolbox{WORC}} framework with the same default configuration on the twelve different datasets are shown in \autoref{fig: AUCBoxplot}; detailed performances metrics are shown in \autoref{tab: allres}; ROC curves are shown in \autoref{fig: ROC} \footnote{Discrepancies between \autoref{tab: allres} and \autoref{tab: rsens} on $N_{\text{RS}}=1000$ and $N_{\text{ens}} = 100$ ensembling are caused by the difference in cross-validation iterations ($k_{\text{test}}=100$ and $k_{\text{test}}=20$, respectively).}. In eleven of the twelve datasets, we successfully found a prediction model; in the Melanoma dataset, the mean AUC was similar to that of guessing (0.50).
	
	\section{Discussion} \label{sec: disc}
	
	For new clinical applications, finding a suitable radiomics workflow out of the wide variety of methods currently has to be done manually, which has many disadvantages. In this study, we proposed the {\toolbox{WORC}} algorithm to automatically construct and optimize radiomics workflows. To this end, we standardized the radiomics workflow in modular components and designed a search space by including a large collection of commonly used radiomics methods and their hyperparameters for each component. We formulated finding the optimal workflow within this search space as a combined model selection and hyperparameter optimization problem and compared two automated machine learning optimization strategies, random search and Bayesian optimization, to solve it. To improve both performance and stability, we proposed various ensembling approaches. To evaluate its performance and generalization, we applied our method to twelve different, independent clinical applications. The results showed that a relatively efficient random search and straight-forward ensembling approach of averaging the best 100 workflows performed at least similar to more advanced methods. While using the exact same configuration, our approach found predictive classification models in eleven applications, substantially outperforming a state-of-the-art radiomics baseline. Hence, our {\toolbox{WORC}} framework automatically streamlines, optimizes, and generalizes radiomics model construction across clinical applications.
	
	In the field of radiomics, there is a lack of reproducibility, while this is vital for the transition of radiomics models to clinical practice \cite{Li2022, Bera2022, RN686, RN436}. A recent study warned that radiomics research must achieve ``higher evidence levels to avoid a reproducibility crisis such as the recent one in psychology'' \cite{RN1089}. Our framework replaces the heuristic trial-and-error process of manual radiomics workflow construction by a fully automated approach, drastically reducing the number of subjective choices and facilitating reproducibility. 
	
	From the twelve datasets in this study, only for the Melanoma dataset was the {\toolbox{WORC}} algorithm not able to find a biomarker. This is studied in detail by \cite{RN791}, which showed that scoring by a radiologist also led to a negative result. This validates our framework, showing that it does not invent a relation when one does not exist. In this way, we hope to contribute to overcoming the challenges for publishing negative results in radiomics \cite{RN696}. 
	
	In seven of the other eleven datasets (Lipo, Desmoid, Liver, GIST, MesFib, Prostate, Glioma), the images were previously visually scored by one or several clinicians, see \autoref{tab: ScoreClinicians} for the scores and \cite{RN574,RN678,starmans2024BLT,RN1246,RN923,RN480,RN492} for details. For five studies (Lipo, Desmoid, Liver, GIST, Prostate), the clinicians` performance was evaluated on a subset or different dataset than the current study and the clinicians were given additional information (e.g., age, sex) besides imaging. In two studies (Lipo, Prostate), the mean of the {\toolbox{WORC}} framework was substantially higher than that of the clinicians, and similar in the other five. In six studies (Lipo, Desmoid, GIST, MesFib, Prostate, Glioma), the {\toolbox{WORC}} framework outperformed at least one of the clinicians. This indicates that the {\toolbox{WORC}} framework has a competitive performance.
	
	There is a trade-off between optimization versus using prior (domain) knowledge to develop a ``logical'' algorithm. For the validation performance, which both {\toolbox SMAC} and the random search optimize, {\toolbox SMAC} substantially outperformed the random search when using a high computational budget. This did however not translate in a higher generalization (i.e., test) performance, indicating method overfitting of {\toolbox SMAC} on the validation dataset. Similar patterns of validation performance overfitting was observed for the more advanced FitNumber and ForwardSelection ensembling approaches compared to the fitting-free $N_{\text{ens}} = 100$ approach. These results led to our proposed choices for random search and $N_{\text{ens}} = 100$ ensembling. Although other optimization or ensembling strategies may improve validation optimization, these results clearly show the limits of optimization and the importance of generalization. Nonetheless, deciding purely based on prior knowledge which algorithm will be optimal is complex and generally not feasible, as the poor performance of our radiomics baseline showed. We suggest to use domain knowledge to determine which algorithms \textit{a priori} have a (near) zero chance of succeeding, e.g., our light fingerprinting approach, while the {\toolbox{WORC}} algorithm optimizes workflow construction within the remaining search space. Lastly, these results also show the risks of method selection overfitting, and thus those of (manual) comparison of different methods on the test dataset, rather warranting automatic method comparison and optimization on the training dataset as in our proposed framework.
	
	Medical deep learning faces similar challenges to conventional radiomics \cite{RN798, RN799, RN800, RN35}, e.g., lack of standardization, wide variety of algorithms, and the need for model selection and hyperparameter tuning per application. Future research may include a similar framework to {\toolbox{WORC}} for deep learning, including the full process from image to prediction, or a hybrid approach. In the field of computer science, automatic deep learning model selection is addressed in Neural Architecture Search (NAS) \cite{RN91}, a hot AutoML topic \cite{RN1106}. In medical imaging, NAS is still at an early stage, with available algorithms mostly focusing on segmentation and optimization of a specific network topology \cite{RN801, Yu_2020_CVPR,VOHO2023369}. While the main concept of our framework, i.e., CASH, could be applied in a similar fashion for deep learning, this poses several challenges. First, deep learning models generally take substantially longer to train, i.e., hours or days compared to less than a second for conventional machine learning methods. Our extensive optimization and cross-validation setup is therefore not feasible. Second, the deep learning search space is less clear due to the wide variety of design choices, while conventional radiomics workflows typically follow the same steps. Lastly, while current NAS approaches focus on architectural design, pre- and post-processing choices may be equally important to include in the search space \cite{RN714, deraad2021preprocessing}. As the pre- and post-processing are performed outside of the network and require \textit{selector} type hyperparameters, combined optimization with the architectural design choices is not trivial.
	
	For the three previously publicly released datasets from other studies (Glioma, Alzheimer, H\&N), we compared the performance of the {\toolbox{WORC}} framework to that of the original studies. In the Glioma dataset, our performance (AUC of 0.72) was the same as the original study (\cite{RN492}: AUC of 0.72). We thus showed that our framework was able to successfully construct a signature using externally precomputed features, and similar to \cite{starmans2024BLT} verified the external-validation setup (\autoref{fig: crossval} b). In the Alzheimer dataset, our performance (AUC of 0.87) was also similar to the original study (\cite{RN1103}: AUC range of 0.80 - 0.94, depending on the level of preprocessing), while we used only hippocampus radiomics features compared to whole-brain voxel-wise features \cite{RN1103}. On the H\&N dataset, we cannot directly compare the results to the original study \cite{RN681}, as the authors did not evaluate the prognostic value of radiomics for predicting the T-stage, and trained a model on a separate dataset of lung tumor patients which is not publicly available anymore. Concluding, to the extent possible when comparing results, the {\toolbox{WORC}} framework showed a similar performance as the original studies.
	
	While the proposed search space includes a large number of commonly used radiomics algorithms, it remains an arbitrary selection: there will always be methods missing. We have formulated the optimization algorithm such that other algorithms and hyperparameters can be added in a straight-forward manner. This facilitates systematic comparison of the new and already included methods, and combining new methods with (parts of) the existing ones to increase overall performance. Hence, when the optimal workflow is not expected to be included in the default {\toolbox{WORC}} search space and a new method is proposed, our framework can be used to systematically evaluate how it can complement existing approaches.
	
	Future research should focus on improving generalization rather than further optimizing validation performance. Firstly, this could include multi-objective optimization \cite{RN962, RN961} to include both validation performance and metrics quantifying generalization, either in the optimization or ensembling. For example, Pareto optimization of both validation accuracy and ensemble diversity can lead to an increased testing performance \cite{RN82}. Secondly, when applying the {\toolbox{WORC}} algorithm on a new dataset, meta-learning could be used to learn from the results on the used twelve datasets \cite{RN694}. Especially on smaller datasets, taking into account which workflows worked best on previous datasets may improve performance, prevent overfitting, and lower computation time. Lastly, our framework may be used on other clinical applications to automatically optimize radiomics workflow construction. While we only showed the use of our framework on CT and MRI, the used features have also been shown to be successful in other modalities such as PET \cite{RN863} and ultrasound \cite{RN864}, and thus the {\toolbox{WORC}} framework could also be useful in these modalities. 

	\section{Conclusions} \label{sec: concl}
	We proposed the {\toolbox{WORC}} framework to fully automatically optimize radiomics workflow construction using automated machine learning and ensembling. The framework was evaluated on twelve different, independent clinical applications, on eleven of which our framework successfully constructed a predictive model. Our framework substantially outperformed a radiomics baseline, and had a competitive performance compared to visual scoring by human experts. Hence, our framework streamlines radiomics research and generalizes radiomics across applications, facilitating systematic data probing for radiomics signatures by automatically comparing and combining thousand radiomics methods. By releasing the {\toolbox{WORC}} database with six datasets of in total 930 patients publicly \cite{WORCDatabase}, and the {\toolbox{WORC}} toolbox implementing our framework plus the code to reproduce the experiments of this study open-source \cite{RN63, WORCDatabasesoftware}, we facilitate radiomics reproducibility and validation.
	
	\section*{Data Statement}
	Six of the datasets used in this study (Lipo, Desmoid, Liver, GIST, CRLM, and Melanoma), comprising a total of 930 patients, are publicly released as part of this study and hosted via a public XNAT\footnote{\href{https://xnat.bmia.nl/data/projects/worc}{https://xnat.bmia.nl/data/projects/worc}} as published in \cite{WORCDatabase}. By storing all data on {\toolbox{XNAT}} in a structured and standardized manner, experiments using these datasets can be easily executed at various computational resources with the same code.
	
	Three datasets were already publicly available as described in \autoref{sec: datasets}. The other three datasets could not be made publicly available. The code for the experiments on the nine publicly available datasets is available on GitHub \cite{WORCDatabasesoftware}.
	
	\section*{Acknowledgments}
	The authors thank Laurens Groenendijk for his assistance in processing the data and in the anonymization procedures, and Hakim Achterberg for his assistance in the development of the software. This work was partially carried out on the Dutch national e-infrastructure with the support of SURF Cooperative.
	
	Data collection and sharing for this project was partially funded by the Alzheimer's Disease Neuroimaging Initiative (ADNI) (National Institutes of Health Grant U01 AG024904) and DOD ADNI (Department of Defense award number W81XWH-12-2-0012). ADNI is funded by the National Institute on Aging, the National Institute of Biomedical Imaging and Bioengineering, and through generous contributions from the following: AbbVie, Alzheimer’s Association; Alzheimer’s Drug Discovery Foundation; Araclon Biotech; BioClinica, Inc.; Biogen; Bristol-Myers Squibb Company; CereSpir, Inc.; Cogstate; Eisai Inc.; Elan Pharmaceuticals, Inc.; Eli Lilly and Company; EuroImmun; F. Hoffmann-La Roche Ltd and its affiliated company Genentech, Inc.; Fujirebio; GE Healthcare; IXICO Ltd.; Janssen Alzheimer Immunotherapy Research \& Development, LLC.; Johnson \& Johnson Pharmaceutical Research \& Development LLC.; Lumosity; Lundbeck; Merck \& Co., Inc.; Meso Scale Diagnostics, LLC.; NeuroRx Research; Neurotrack Technologies; Novartis Pharmaceuticals Corporation; Pfizer Inc.; Piramal Imaging; Servier; Takeda Pharmaceutical Company; and Transition Therapeutics. The Canadian Institutes of Health Research is providing funds to support ADNI clinical sites in Canada. Private sector contributions are facilitated by the Foundation for the National Institutes of Health (\href{www.fnih.org}{www.fnih.org}). The grantee organization is the Northern California Institute for Research and Education, and the study is coordinated by the Alzheimer’s Therapeutic Research Institute at the University of Southern California. ADNI data are disseminated by the Laboratory for Neuro Imaging at the University of Southern California.
	
	\section*{Funding}
	Martijn P. A. Starmans and Jose M. Castillo T. acknowledge funding from the research program STRaTeGy with project numbers 14929, 14930, and 14932, which is (partly) financed by the Netherlands Organization for Scientific Research (NWO). Sebastian R. van der Voort acknowledges funding from the Dutch Cancer Society (KWF project number EMCR 2015-7859). Part of this study was financed by the Stichting Coolsingel (reference number 567), a Dutch non-profit foundation. This study is supported by EuCanShare and EuCanImage (European Union's Horizon 2020 research and innovation programme under grant agreement Nr. 825903 and Nr. 952103, respectively). 
	
	\section*{Competing Interests Statement}
	Wiro J. Niessen is founder, scientific lead, and shareholder of Quantib BV. Jacob J. Visser is a medical advisor at Contextflow. Astrid A. M. van der Veldt is a consultant (fees paid to the institute) at BMS, Merck, MSD, Sanofi, Eisai, Pfizer, Roche, Novartis, Pierre Fabre and Ipsen. The other authors do not declare any conflicts of interest.
	
	\section*{CRediT Author Statement}
	M.P.A.S., W.J.N., and S.K. provided the conception and design of the study. M.P.A.S., M.J.M.T., M.V., G.A.P., W.K., D.H., D.J.G., C.V., S.S., R.S.D., C.J.E., F.F., G.J.L.H.v.L., A.B., J.H, T.B., R.v.G., G.J.H.F., R.A.F., W.W.d.H., F.E.B., F.E.J.A.W., B.G.K., L.A., A.A.M.v.d.V., A.R., A.E.O., J.M.C.T., J.V., I.S., M.R., Mic.D., R.d.M., J.IJ., R.L.M., P.B.V., E.E.B., M.G.T., and J.J.V. acquired the data. M.P.A.S., S.R.v.d.V., M.J.M.T., M.V., A.B., F.E.B., L.A., Mit.D., J.M.C.T., R.L.M., E.B., M.G.T. and S.K. analyzed and interpreted the data. M.P.A.S., S.R.v.d.V., T.P., and Mit.D. created the software. M.P.A.S. and S.K. drafted the article. All authors read and approved the final manuscript.
	
	\section*{Ethics Statement}
	The protocol of this study conformed to the ethical guidelines of the 1975 Declaration of Helsinki. Approval by the local institutional review board of the Erasmus MC (Rotterdam, the Netherlands) was obtained for collection of the WORC database (MEC-2020-0961), and separately for eight of the studies using in-house data (Lipo: MEC-2016-339, Desmoid: MEC-2016-339, Liver: MEC-2017-1035, GIST: MEC-2017-1187, CRLM: MEC-2017-479, Melanoma: MEC-2019-0693, HCC: MEC-2018-1621, Prostate: NL32105.078.10). The need for informed consent was waived due to the use of anonymized, retrospective data. For the last study involving in-house data, the Mesfib study, as the study was retrospectively performed with anonymized data, no approval from the ethical committee or informed consent was required.
	
	\bibliographystyle{IEEEtran}
	\bibliography{EndnoteLibraryMartijn_Parsed}

{
		
		\appendices
		
		\section{Details on included radiomics algorithms and hyperparameters} \label{sec: SupplementaryMaterial1}
		This supplementary material includes details on the radiomics algorithms and their associated hyperparameters included in the default search space of the {\toolbox{WORC}} optimization algorithm. These are discussed per component of the radiomics workflow: image and segmentation preprocessing (\ref{sec: MethodsImSegPreprocess}), feature extraction (\ref{sec: MethodsFeatures}), feature and sample preprocessing (\ref{sec: MethodsFeaturePreprocessing}), and machine learning (\ref{sec: MethodsML}).
		
		
		\subsection{Image and segmentation preprocessing} \label{sec: MethodsImSegPreprocess}
		Before feature extraction, image preprocessing such as image quantization, normalization, resampling or noise filtering may be applied \cite{RN261, RN238, RN761}. By default no preprocessing is applied. The only exception is image normalization (using z-scoring), which we apply in modalities that do not have a fixed unit and scale (e.g. qualitative MRI, ultrasound), but not in modalities that have a fixed unit and scale (e.g. Computed Tomography (CT), quantitative MRI such as T1 mapping). A fingerprinting approach is used to decide between using image normalization or not, see \autoref{sec: Fingerprinting}.
		
		\subsection{Feature extraction} \label{sec: MethodsFeatures}
		For each segmentation, 564 radiomics features quantifying intensity, shape, orientation and texture are extracted through the open-source feature toolboxes {\toolbox PyRadiomics} \cite{RN761} and {\toolbox PREDICT} \cite{RN373}. A comprehensive overview is provided in \autoref{tab: Features}. Thirteen intensity features describe various first-order statistics of the raw intensity distributions within the segmentation, such as the mean, standard deviation, and kurtosis. Thirty-five shape features describe the morphological properties of the segmentation, and are extracted based only on the segmentation, i.e., not using the image. These include shape descriptions such as the volume, compactness, and circular variance. Nine orientation features describe the orientation and positioning of the segmentation, i.e., not using the image. These include the major axis orientations of a 3D ellipse fitted to the segmentation, the center of mass coordinates and indices. Lastly, 507 texture features are extracted, which include commonly used algorithms such as the Gray Level Co-occurence Matrix (GLCM) (144 features) \cite{RN539}, Gray Level Size Zone Matrix (GLSZM) (16 features) \cite{RN539}, Gray Level Run Length Matrix (GLRLM) (16 features) \cite{RN539}, Gray Level Dependence Matrix (GLDM) (14 features) \cite{RN539}, Neighborhood Grey Tone Difference Matrix (NGTDM) (5 features) \cite{RN539}, Gabor filters (156 features) \cite{RN539}, Laplacian of Gaussian (LoG) filters (39 features) \cite{RN539}, and Local Binary Patterns (LBP) (39 features) \cite{RN270}. Additionally, two less common feature groups are defined: based on local phase \cite{RN140} (39 features) and vesselness filters \cite{RN583} (39 features).
		
		Many radiomics studies include datasets with variations in the slice thickness due to heterogeneity in the acquisition protocols. This may cause feature values to be dependent on the acquisition protocol. Moreover, the slice thickness is often substantially larger than the pixel spacing. Hence, extracting robust 3D features may be hampered by these variations, especially for low resolutions. To overcome this issue, a 2.5D approach can be used:  features are extracted per 2D axial slice and aggregated over all slices. Afterwards, several first-order statistics over the feature distributions are evaluated and used as actual features, see also \autoref{tab: Features}. A fingerprinting approach is currently used to decide between 2D, 2.5D, and 3D feature extraction, see \autoref{sec: Fingerprinting}.
		
		Some of the features have parameters themselves, such as the scale on which a derivative is taken. As some features are rather computationally expensive to extract, we do not include these parameters directly as hyperparameters in the CASH problem. Instead, the features are extracted for a predefined range of parameter values. In the next components, feature selection algorithms are employed to select the most relevant features and thus parameters. The used parameter ranges are reported in \autoref{tab: Features}.
		
		Radiomics studies may involve multiple scans per sample, e.g. in multimodal (MRI + CT) or multi-contrast (T1-weighted MRI + T2-weighted MRI) studies. Commonly, radiomics features are defined on a single image, which also holds for the features described in this study. Hence, when multiple scans per sample are included, the 564 radiomics features are extracted per scan and concatenated, which in this study only is used in the Prostate dataset.
		
		\subsection{Feature and sample preprocessing} \label{sec: MethodsFeaturePreprocessing}
		We define feature and sample preprocessing as all algorithms that can be used between the feature extraction and machine learning components. The order of these algorithms in the {\toolbox{WORC}} framework is fixed and given in \autoref{table: hyperparameters}.
		
		Feature imputation is employed to replace missing feature values. Values may be missing when a feature could not be defined and computed, e.g. a lesion may be too small for a specific feature to be extracted. Algorithms for imputation include: 1) mean; 2) median; 3) mode; 4) constant value (default: zero); and 5) nearest neighbor approach.
		
		Feature scaling is employed to ensure that all features have a similar scale. As this generally benefits machine learning algorithms, this is always performed through z-scoring. A robust version is used, where outliers, defined as feature values outside the $5^{\text{th}} - 95^{\text{th}}$ percentile range are excluded before computation of the mean and standard deviation.
		
		Feature selection or dimensionality reduction algorithms may be employed to select the most relevant features and eliminate irrelevant or redundant features. As multiple algorithms may be combined, instead of defining feature selection or dimensionality reduction as a single step, each algorithm is included as a single step in the workflow with an \textit{activator} hyperparameter to determine whether the algorithm is used or not.
		
		Algorithms included are:
		\begin{enumerate}
			\item A group-wise feature selection, in which groups of features (i.e., intensity, shape, and texture feature subgroups) can be selected or eliminated. To this end, each feature group has an \textit{activator} hyperparameter. This algorithm serves as regularization, as it randomly reduces the feature set, and is therefore always used. The group-wise feature selection is the first step in the workflows, as it reduces the computation time of the other steps by reducing the feature space.
			\item A variance threshold, in which features with a low variance ($<0.01$) are removed. This algorithm is always used, as this serves as a feature sanity check with almost zero risk of removing relevant features. The variance threshold is applied before the feature scaling, as this results in all features having unit variance.
			\item Optionally, the RELIEF algorithm \cite{RN513}, which ranks the features according to the differences between neighboring samples. Features with more differences between neighbors of different classes are considered higher in rank.
			\item Optionally, feature selection using a machine learning model \cite{RN514}. Features are selected based on their importance as given by a machine learning model trained on the dataset. Hence, the used algorithm should be able to give the features an importance weight. Algorithms included are LASSO, logistic regression, and random forest.
			\item Optionally, principal component analysis (PCA), in which either only those linear combinations of features are kept which explained 95\% of the variance in the features, or a fixed number of components (10, 50, or 100) is selected.
			\item Optionally, individual feature selection through univariate testing. To this end, for each feature, a Mann-Whitney U test is performed to test for significant differences in distribution between the classes. Afterwards, only features with p-values below a certain threshold are selected. The (non-parametric) Mann-Whitney U test was chosen as it makes no assumptions about the distribution of the features.
		\end{enumerate}
		
		RELIEF, selection using a model, PCA, and univariate testing have a 27.5\% chance to be included in a workflow in the random search, as this gives an equal chance of applying any of these or no feature selection algorithm. The feature selection algorithms may only be combined in the mentioned order in the {\toolbox{WORC}} framework.
		
		Resampling algorithms may be used, primarily to deal with class imbalances. These include various algorithms from the {\toolbox{imbalanced-learn}} toolbox \cite{RN814}: 1) random under-sampling; 2) random over-sampling; 3) near-miss resampling; 4) the neighborhood cleaning rule; 5) SMOTE \cite{RN677} (regular, borderline, Tomek, and the edited nearest neighbors variant); and 6) ADASYN \cite{RN703}. All algorithms can apply four out of five different resampling strategies, resampling: 1) the minority class (not for undersampling algorithms); 2) all but the minority class; 3) the majority class (not for oversampling algorithms); 4) all but the majority class; and 5) all classes.
		
		\subsection{Machine learning} \label{sec: MethodsML}
		For machine learning, we mostly use methods from the {\toolbox{scikit-learn}} toolbox \cite{RN516}. The following classification algorithms are included: 1) logistic regression; 2) support vector machines (with a linear, polynomial, or radial basis function kernel); 3) random forests; 4) naive Bayes; 5) linear discriminant analysis; 6) quadratic discriminant analysis (QDA); 7) AdaBoost \cite{RN698}; and 8) extreme gradient boosting (XGBoost) \cite{RN700}. The associated hyperparameters for each algorithm are depicted in \autoref{table: hyperparameters}.
		
		\section{Hardware and computation time}\label{sec: SupplementaryMaterial2Computationtime}
		The computation time of a {\toolbox{WORC}} experiment roughly scales with $k_{\text{training}}$, $k_{\text{test}}$, and $N_{RS}$ for the random search or the {\toolbox SMAC} computational budget. A high degree of parallelization for all these parameters is possible, as all workflows can be independently executed. We choose to run the iterations of $k_{\text{test}}$ sequentially instead of in parallel to maintain a sustainable computational load. For the $k_{\text{training}}$ iterations and $N_{RS}$ samples, all workflows are run in parallel. An experiment consisting of executing 500000 workflows ($k_{\text{training}} = 5$, $k_{\text{test}} = 100$, and $N_{RS} = 1000$) on average had a computation time of approximately 18 hours on a machine with 24 Intel E5-2695 v2 CPU cores, hence roughly 10 minutes per train-test cross-validation iteration. The contribution of the feature extraction to the computation time is negligible.
		
	\newpage
	
	\renewcommand{\thefigure}{A.\arabic{figure}}
	\setcounter{figure}{0}
	\renewcommand{\thetable}{A.\arabic{table}}
	\setcounter{table}{0}
	\renewcommand{\thealgorithm}{A.\arabic{algorithm}}
	\setcounter{algorithm}{0}  
	
	\begin{figure*}
		\centering
		\includegraphics[width=0.9\textwidth]{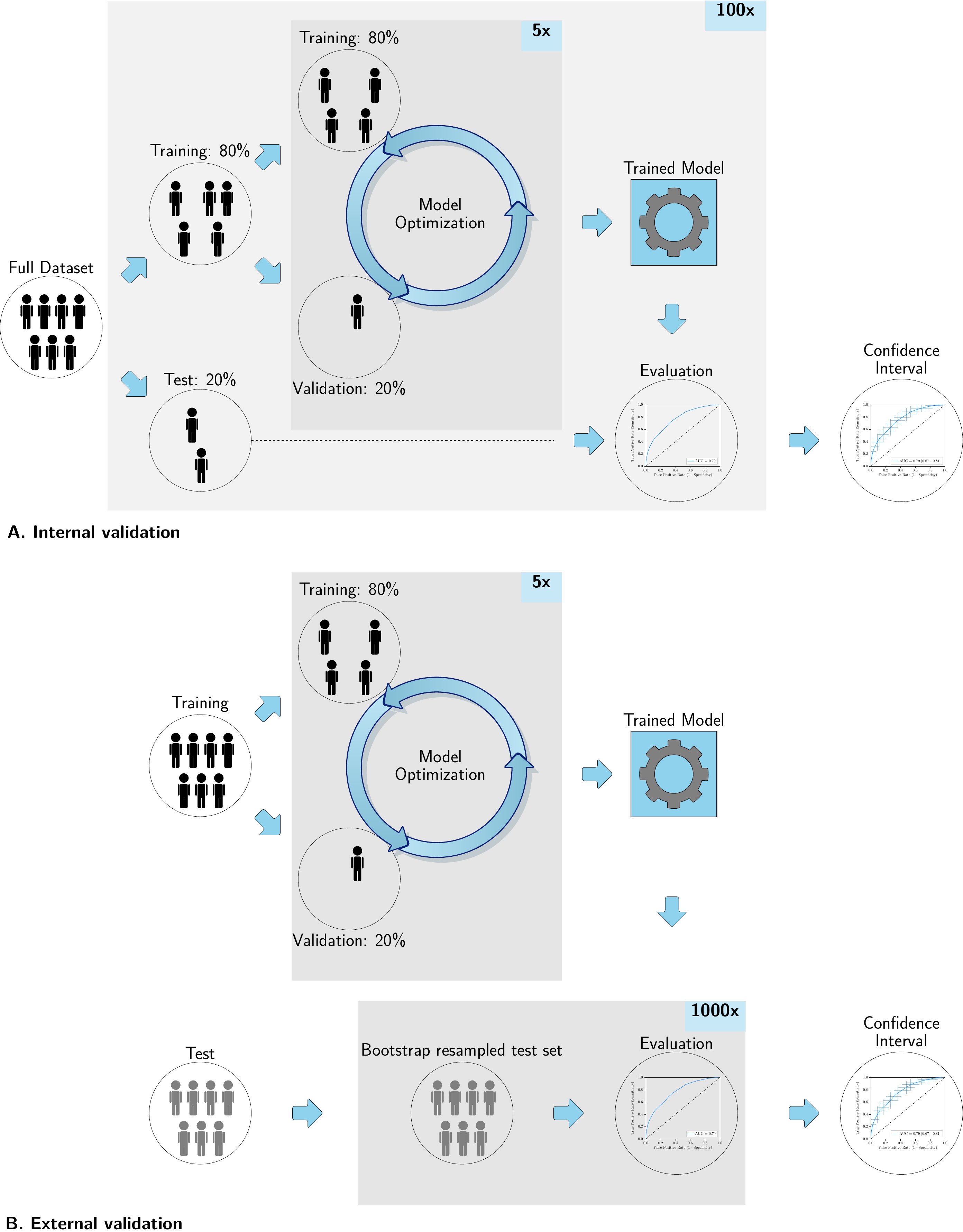}
		\caption{Cross-validation setups used by our {\toolbox{WORC}} framework for optimization and evaluation. When a single dataset is used, internal validation is performed through a $k_{\text{test}}=100$ random-split cross-validation (\textbf{A}). When fixed, separate training and test datasets are used, external validation is performed by developing the model on the training set and evaluating the performance on the test set through 1000x bootstrap resampling (\textbf{B}). Both include an internal $k_{\text{training}}=5$ random-split cross-validation on the training set to split the training set into parts for actual training and validation, in which the model optimization is performed. The final selected model, trained on the full training dataset, is used for independent testing on the test dataset.}
		\label{fig: crossval}
	\end{figure*}
	
	\begin{table*}
		\centering
		\resizebox{0.75\textwidth}{!}{
			\begin{tabular}{lllllllllll}
				\textbf{Lipo} &
				\multicolumn{2}{l}{$N_{\text{RS}}   = 10$} &
				\multicolumn{2}{l}{$N_{\text{RS}}   = 100$} &
				\multicolumn{2}{l}{$N_{\text{RS}}   = 1000$} &
				\multicolumn{2}{l}{$N_{\text{RS}}   = 10000$} &
				\multicolumn{2}{l}{$N_{\text{RS}}   = 25000$} \\
				&
				\textbf{Mean} &
				\textbf{Std} &
				\textbf{Mean} &
				\textbf{Std} &
				\textbf{Mean} &
				\textbf{Std} &
				\textbf{Mean} &
				\textbf{Std} &
				\textbf{Mean} &
				\\
				$N_{\text{ens}} = 1$ &
				\cellcolor[HTML]{FBA776}0.69 &
				\cellcolor[HTML]{A4A4A4}0.024 &
				\cellcolor[HTML]{EBE683}0.73 &
				\cellcolor[HTML]{C7C7C7}0.016 &
				\cellcolor[HTML]{FEE382}0.73 &
				\cellcolor[HTML]{C7C7C7}0.016 &
				\cellcolor[HTML]{FEDD81}0.72 &
				\cellcolor[HTML]{DCDCDC}0.011 &
				\cellcolor[HTML]{FDCA7D}0.71 &
				\\
				$N_{\text{ens}} = 10$ &
				\cellcolor[HTML]{FA9272}0.68 &
				\cellcolor[HTML]{CBCBCB}0.015 &
				\cellcolor[HTML]{F1E784}0.73 &
				\cellcolor[HTML]{DCDCDC}0.011 &
				\cellcolor[HTML]{98CE7F}0.75 &
				\cellcolor[HTML]{D8D8D8}0.012 &
				\cellcolor[HTML]{98CE7F}0.75 &
				\cellcolor[HTML]{EEEEEE}0.007 &
				\cellcolor[HTML]{8DCA7E}0.75 &
				\\
				$N_{\text{ens}} = 100$ &
				- &
				- &
				\cellcolor[HTML]{FDD37F}0.72 &
				\cellcolor[HTML]{EAEAEA}0.008 &
				\cellcolor[HTML]{F1E784}0.73 &
				\cellcolor[HTML]{F7F7F7}0.005 &
				\cellcolor[HTML]{81C77D}0.75 &
				\cellcolor[HTML]{F7F7F7}0.005 &
				\cellcolor[HTML]{98CE7F}0.75 &
				\\
				FitNumber &
				\cellcolor[HTML]{F8696B}0.65 &
				\cellcolor[HTML]{ACACAC}0.022 &
				\cellcolor[HTML]{FCB77A}0.70 &
				\cellcolor[HTML]{CFCFCF}0.014 &
				\cellcolor[HTML]{FEE783}0.73 &
				\cellcolor[HTML]{DCDCDC}0.011 &
				\cellcolor[HTML]{D3DF82}0.74 &
				\cellcolor[HTML]{C2C2C2}0.017 &
				\cellcolor[HTML]{FEE783}0.73 &
				\\
				ForwardSelection &
				\cellcolor[HTML]{F98470}0.67 &
				\cellcolor[HTML]{D4D4D4}0.013 &
				\cellcolor[HTML]{FDD27F}0.72 &
				\cellcolor[HTML]{EEEEEE}0.007 &
				\cellcolor[HTML]{AAD380}0.75 &
				\cellcolor[HTML]{F7F7F7}0.005 &
				\cellcolor[HTML]{93CC7E}0.75 &
				\cellcolor[HTML]{EEEEEE}0.007 &
				\cellcolor[HTML]{63BE7B}0.76 &
				\\
				&
				&
				&
				&
				&
				&
				&
				&
				&
				&
				\\
				\textbf{Desmoid} &
				\multicolumn{2}{l}{$N_{\text{RS}} = 10$} &
				\multicolumn{2}{l}{$N_{\text{RS}} = 100$} &
				\multicolumn{2}{l}{$N_{\text{RS}} = 1000$} &
				\multicolumn{2}{l}{$N_{\text{RS}} = 10000$} &
				\multicolumn{2}{l}{$N_{\text{RS}} = 25000$} \\
				&
				\textbf{Mean} &
				\textbf{Std} &
				\textbf{Mean} &
				\textbf{Std} &
				\textbf{Mean} &
				\textbf{Std} &
				\textbf{Mean} &
				\textbf{Std} &
				\textbf{Mean} &
				\\
				$N_{\text{ens}} = 1$ &
				\cellcolor[HTML]{FA9D75}0.70 &
				\cellcolor[HTML]{BEBEBE}0.018 &
				\cellcolor[HTML]{FCB97A}0.72 &
				\cellcolor[HTML]{EEEEEE}0.007 &
				\cellcolor[HTML]{FCC07B}0.72 &
				\cellcolor[HTML]{DCDCDC}0.011 &
				\cellcolor[HTML]{FCC37C}0.73 &
				\cellcolor[HTML]{E5E5E5}0.009 &
				\cellcolor[HTML]{FED880}0.74 &
				\\
				$N_{\text{ens}} = 10$ &
				\cellcolor[HTML]{FCB679}0.72 &
				\cellcolor[HTML]{CFCFCF}0.014 &
				\cellcolor[HTML]{FFEB84}0.75 &
				\cellcolor[HTML]{EEEEEE}0.007 &
				\cellcolor[HTML]{CCDD82}0.76 &
				\cellcolor[HTML]{EEEEEE}0.007 &
				\cellcolor[HTML]{FEE382}0.75 &
				\cellcolor[HTML]{EEEEEE}0.007 &
				\cellcolor[HTML]{FFEB84}0.75 &
				\\
				$N_{\text{ens}} = 100$ &
				- &
				- &
				\cellcolor[HTML]{FFEB84}0.75 &
				\cellcolor[HTML]{F2F2F2}0.006 &
				\cellcolor[HTML]{63BE7B}0.76 &
				\cellcolor[HTML]{FBFBFB}0.004 &
				\cellcolor[HTML]{CCDD82}0.76 &
				\cellcolor[HTML]{FFFFFF}0.003 &
				\cellcolor[HTML]{98CE7F}0.76 &
				\\
				FitNumber &
				\cellcolor[HTML]{F8696B}0.66 &
				\cellcolor[HTML]{CBCBCB}0.015 &
				\cellcolor[HTML]{FDD17F}0.74 &
				\cellcolor[HTML]{CBCBCB}0.015 &
				\cellcolor[HTML]{FFEB84}0.75 &
				\cellcolor[HTML]{EAEAEA}0.008 &
				\cellcolor[HTML]{FEE382}0.75 &
				\cellcolor[HTML]{EEEEEE}0.007 &
				\cellcolor[HTML]{75C37C}0.76 &
				\\
				ForwardSelection &
				\cellcolor[HTML]{FAA075}0.70 &
				\cellcolor[HTML]{D8D8D8}0.012 &
				\cellcolor[HTML]{86C97E}0.76 &
				\cellcolor[HTML]{F2F2F2}0.006 &
				\cellcolor[HTML]{75C37C}0.76 &
				\cellcolor[HTML]{F7F7F7}0.005 &
				\cellcolor[HTML]{CCDD82}0.76 &
				\cellcolor[HTML]{F2F2F2}0.006 &
				\cellcolor[HTML]{DDE182}0.76 &
				\\
				&
				&
				&
				&
				&
				&
				&
				&
				&
				&
				\\
				\textbf{Liver} &
				\multicolumn{2}{l}{$N_{\text{RS}} = 10$} &
				\multicolumn{2}{l}{$N_{\text{RS}} = 100$} &
				\multicolumn{2}{l}{$N_{\text{RS}} = 1000$} &
				\multicolumn{2}{l}{$N_{\text{RS}} = 10000$} &
				\multicolumn{2}{l}{$N_{\text{RS}} = 25000$} \\
				&
				\textbf{Mean} &
				\textbf{Std} &
				\textbf{Mean} &
				\textbf{Std} &
				\textbf{Mean} &
				\textbf{Std} &
				\textbf{Mean} &
				\textbf{Std} &
				\textbf{Mean} &
				\\
				$N_{\text{ens}} = 1$ &
				\cellcolor[HTML]{F98871}0.64 &
				\cellcolor[HTML]{C2C2C2}0.017 &
				\cellcolor[HTML]{FCC27C}0.68 &
				\cellcolor[HTML]{CFCFCF}0.014 &
				\cellcolor[HTML]{FDC97D}0.68 &
				\cellcolor[HTML]{BEBEBE}0.018 &
				\cellcolor[HTML]{FDC87D}0.68 &
				\cellcolor[HTML]{E5E5E5}0.009 &
				\cellcolor[HTML]{FDCE7E}0.69 &
				\\
				$N_{\text{ens}} = 10$ &
				\cellcolor[HTML]{FA9F75}0.66 &
				\cellcolor[HTML]{C7C7C7}0.016 &
				\cellcolor[HTML]{CFDE82}0.71 &
				\cellcolor[HTML]{F7F7F7}0.005 &
				\cellcolor[HTML]{88C97E}0.72 &
				\cellcolor[HTML]{EEEEEE}0.007 &
				\cellcolor[HTML]{FEE582}0.70 &
				\cellcolor[HTML]{F7F7F7}0.005 &
				\cellcolor[HTML]{E8E583}0.71 &
				\\
				$N_{\text{ens}} = 100$ &
				- &
				- &
				\cellcolor[HTML]{F4E884}0.71 &
				\cellcolor[HTML]{EAEAEA}0.008 &
				\cellcolor[HTML]{63BE7B}0.72 &
				\cellcolor[HTML]{F7F7F7}0.005 &
				\cellcolor[HTML]{C4DA81}0.71 &
				\cellcolor[HTML]{FBFBFB}0.004 &
				\cellcolor[HTML]{F4E884}0.71 &
				\\
				FitNumber &
				\cellcolor[HTML]{F8696B}0.62 &
				\cellcolor[HTML]{B5B5B5}0.020 &
				\cellcolor[HTML]{FDD27F}0.69 &
				\cellcolor[HTML]{CFCFCF}0.014 &
				\cellcolor[HTML]{93CC7E}0.72 &
				\cellcolor[HTML]{F2F2F2}0.006 &
				\cellcolor[HTML]{B7D780}0.71 &
				\cellcolor[HTML]{EEEEEE}0.007 &
				\cellcolor[HTML]{FEE883}0.71 &
				\\
				ForwardSelection &
				\cellcolor[HTML]{FA9D75}0.65 &
				\cellcolor[HTML]{B1B1B1}0.021 &
				\cellcolor[HTML]{B7D780}0.71 &
				\cellcolor[HTML]{DCDCDC}0.011 &
				\cellcolor[HTML]{63BE7B}0.72 &
				\cellcolor[HTML]{F2F2F2}0.006 &
				\cellcolor[HTML]{E8E583}0.71 &
				\cellcolor[HTML]{FFFFFF}0.003 &
				\cellcolor[HTML]{FEE983}0.71 &
				\\
				&
				&
				&
				&
				&
				&
				&
				&
				&
				&
				\\
				\textbf{GIST} &
				\multicolumn{2}{l}{$N_{\text{RS}} = 10$} &
				\multicolumn{2}{l}{$N_{\text{RS}} = 100$} &
				\multicolumn{2}{l}{$N_{\text{RS}} = 1000$} &
				\multicolumn{2}{l}{$N_{\text{RS}} = 10000$} &
				\multicolumn{2}{l}{$N_{\text{RS}} = 25000$} \\
				&
				\textbf{Mean} &
				\textbf{Std} &
				\textbf{Mean} &
				\textbf{Std} &
				\textbf{Mean} &
				\textbf{Std} &
				\textbf{Mean} &
				\textbf{Std} &
				\textbf{Mean} &
				\\
				$N_{\text{ens}} = 1$ &
				\cellcolor[HTML]{FA9B74}0.63 &
				\cellcolor[HTML]{E1E1E1}0.010 &
				\cellcolor[HTML]{FCC17B}0.66 &
				\cellcolor[HTML]{D8D8D8}0.012 &
				\cellcolor[HTML]{FDD780}0.67 &
				\cellcolor[HTML]{E1E1E1}0.010 &
				\cellcolor[HTML]{FDCE7E}0.67 &
				\cellcolor[HTML]{D8D8D8}0.012 &
				\cellcolor[HTML]{FCBC7B}0.65 &
				\\
				$N_{\text{ens}} = 10$ &
				\cellcolor[HTML]{FA9673}0.63 &
				\cellcolor[HTML]{C7C7C7}0.016 &
				\cellcolor[HTML]{F6E984}0.69 &
				\cellcolor[HTML]{EEEEEE}0.007 &
				\cellcolor[HTML]{AAD380}0.70 &
				\cellcolor[HTML]{EEEEEE}0.007 &
				\cellcolor[HTML]{C3DA81}0.70 &
				\cellcolor[HTML]{EEEEEE}0.007 &
				\cellcolor[HTML]{FEE883}0.69 &
				\\
				$N_{\text{ens}} = 100$ &
				- &
				- &
				\cellcolor[HTML]{FDC97D}0.66 &
				\cellcolor[HTML]{EAEAEA}0.008 &
				\cellcolor[HTML]{63BE7B}0.71 &
				\cellcolor[HTML]{F7F7F7}0.005 &
				\cellcolor[HTML]{70C27C}0.71 &
				\cellcolor[HTML]{FFFFFF}0.003 &
				\cellcolor[HTML]{77C47D}0.71 &
				\\
				FitNumber &
				\cellcolor[HTML]{F8696B}0.59 &
				\cellcolor[HTML]{9B9B9B}0.026 &
				\cellcolor[HTML]{FBA977}0.64 &
				\cellcolor[HTML]{CFCFCF}0.014 &
				\cellcolor[HTML]{D0DE82}0.70 &
				\cellcolor[HTML]{EAEAEA}0.008 &
				\cellcolor[HTML]{9DCF7F}0.70 &
				\cellcolor[HTML]{EEEEEE}0.007 &
				\cellcolor[HTML]{F0E784}0.69 &
				\\
				ForwardSelection &
				\cellcolor[HTML]{FBA175}0.63 &
				\cellcolor[HTML]{CFCFCF}0.014 &
				\cellcolor[HTML]{FEE382}0.68 &
				\cellcolor[HTML]{E5E5E5}0.009 &
				\cellcolor[HTML]{77C47D}0.71 &
				\cellcolor[HTML]{F2F2F2}0.006 &
				\cellcolor[HTML]{8ACA7E}0.71 &
				\cellcolor[HTML]{F2F2F2}0.006 &
				\cellcolor[HTML]{9DCF7F}0.70 &
				\\
				&
				&
				&
				&
				&
				&
				&
				&
				&
				&
				\\
				\textbf{CRLM} &
				\multicolumn{2}{l}{$N_{\text{RS}} = 10$} &
				\multicolumn{2}{l}{$N_{\text{RS}} = 100$} &
				\multicolumn{2}{l}{$N_{\text{RS}} = 1000$} &
				\multicolumn{2}{l}{$N_{\text{RS}} = 10000$} &
				\multicolumn{2}{l}{$N_{\text{RS}} = 25000$} \\
				&
				\textbf{Mean} &
				\textbf{Std} &
				\textbf{Mean} &
				\textbf{Std} &
				\textbf{Mean} &
				\textbf{Std} &
				\textbf{Mean} &
				\textbf{Std} &
				\textbf{Mean} &
				\\
				$N_{\text{ens}} = 1$ &
				\cellcolor[HTML]{F8696B}0.53 &
				\cellcolor[HTML]{ACACAC}0.022 &
				\cellcolor[HTML]{F86E6B}0.54 &
				\cellcolor[HTML]{CFCFCF}0.014 &
				\cellcolor[HTML]{FCC47C}0.55 &
				\cellcolor[HTML]{969696}0.027 &
				\cellcolor[HTML]{FCB579}0.55 &
				\cellcolor[HTML]{ACACAC}0.022 &
				\cellcolor[HTML]{F6E984}0.56 &
				\\
				$N_{\text{ens}} = 10$ &
				\cellcolor[HTML]{FCBF7B}0.55 &
				\cellcolor[HTML]{858585}0.031 &
				\cellcolor[HTML]{FDCE7E}0.55 &
				\cellcolor[HTML]{CFCFCF}0.014 &
				\cellcolor[HTML]{FDD47F}0.56 &
				\cellcolor[HTML]{C2C2C2}0.017 &
				\cellcolor[HTML]{BED981}0.58 &
				\cellcolor[HTML]{CFCFCF}0.014 &
				\cellcolor[HTML]{63BE7B}0.60 &
				\\
				$N_{\text{ens}} = 100$ &
				- &
				- &
				\cellcolor[HTML]{85C87D}0.59 &
				\cellcolor[HTML]{E1E1E1}0.010 &
				\cellcolor[HTML]{F6E984}0.56 &
				\cellcolor[HTML]{D8D8D8}0.012 &
				\cellcolor[HTML]{D8E082}0.57 &
				\cellcolor[HTML]{E5E5E5}0.009 &
				\cellcolor[HTML]{DCE182}0.57 &
				\\
				FitNumber &
				\cellcolor[HTML]{F8786D}0.54 &
				\cellcolor[HTML]{808080}0.032 &
				\cellcolor[HTML]{E3E383}0.57 &
				\cellcolor[HTML]{BEBEBE}0.018 &
				\cellcolor[HTML]{FBAB77}0.55 &
				\cellcolor[HTML]{B5B5B5}0.020 &
				\cellcolor[HTML]{FEDE81}0.56 &
				\cellcolor[HTML]{CFCFCF}0.014 &
				\cellcolor[HTML]{FBA175}0.55 &
				\\
				ForwardSelection &
				\cellcolor[HTML]{FBB078}0.55 &
				\cellcolor[HTML]{969696}0.027 &
				\cellcolor[HTML]{C1DA81}0.58 &
				\cellcolor[HTML]{D4D4D4}0.013 &
				\cellcolor[HTML]{E7E483}0.57 &
				\cellcolor[HTML]{C2C2C2}0.017 &
				\cellcolor[HTML]{A3D17F}0.58 &
				\cellcolor[HTML]{E5E5E5}0.009 &
				\cellcolor[HTML]{76C47D}0.60 &
				\\
				&
				&
				&
				&
				&
				&
				&
				&
				&
				&
				\\
				\textbf{Melanoma} &
				\multicolumn{2}{l}{$N_{\text{RS}} = 10$} &
				\multicolumn{2}{l}{$N_{\text{RS}} = 100$} &
				\multicolumn{2}{l}{$N_{\text{RS}} = 1000$} &
				\multicolumn{2}{l}{$N_{\text{RS}} = 10000$} &
				\multicolumn{2}{l}{$N_{\text{RS}} = 25000$} \\
				&
				\textbf{Mean} &
				\textbf{Std} &
				\textbf{Mean} &
				\textbf{Std} &
				\textbf{Mean} &
				\textbf{Std} &
				\textbf{Mean} &
				\textbf{Std} &
				\textbf{Mean} &
				\\
				$N_{\text{ens}} = 1$ &
				\cellcolor[HTML]{D1DE82}0.46 &
				\cellcolor[HTML]{C2C2C2}0.017 &
				\cellcolor[HTML]{FA9172}0.44 &
				\cellcolor[HTML]{858585}0.031 &
				\cellcolor[HTML]{F0E784}0.46 &
				\cellcolor[HTML]{A8A8A8}0.023 &
				\cellcolor[HTML]{F8716C}0.43 &
				\cellcolor[HTML]{B9B9B9}0.019 &
				\cellcolor[HTML]{FEE282}0.46 &
				\\
				$N_{\text{ens}} = 10$ &
				\cellcolor[HTML]{FFEB84}0.46 &
				\cellcolor[HTML]{B9B9B9}0.019 &
				\cellcolor[HTML]{FCB579}0.45 &
				\cellcolor[HTML]{A8A8A8}0.023 &
				\cellcolor[HTML]{FCB579}0.45 &
				\cellcolor[HTML]{B1B1B1}0.021 &
				\cellcolor[HTML]{FFEB84}0.46 &
				\cellcolor[HTML]{B9B9B9}0.019 &
				\cellcolor[HTML]{73C37C}0.48 &
				\\
				$N_{\text{ens}} = 100$ &
				- &
				- &
				\cellcolor[HTML]{F98370}0.44 &
				\cellcolor[HTML]{E5E5E5}0.009 &
				\cellcolor[HTML]{FA9E75}0.44 &
				\cellcolor[HTML]{F2F2F2}0.006 &
				\cellcolor[HTML]{FDD47F}0.45 &
				\cellcolor[HTML]{EAEAEA}0.008 &
				\cellcolor[HTML]{63BE7B}0.48 &
				\\
				FitNumber &
				\cellcolor[HTML]{FCB579}0.45 &
				\cellcolor[HTML]{CBCBCB}0.015 &
				\cellcolor[HTML]{F97F6F}0.43 &
				\cellcolor[HTML]{BEBEBE}0.018 &
				\cellcolor[HTML]{FFEB84}0.46 &
				\cellcolor[HTML]{ACACAC}0.022 &
				\cellcolor[HTML]{FFEB84}0.46 &
				\cellcolor[HTML]{CBCBCB}0.015 &
				\cellcolor[HTML]{B9D780}0.47 &
				\\
				ForwardSelection &
				\cellcolor[HTML]{FFEB84}0.46 &
				\cellcolor[HTML]{D4D4D4}0.013 &
				\cellcolor[HTML]{F8696B}0.43 &
				\cellcolor[HTML]{D4D4D4}0.013 &
				\cellcolor[HTML]{F0E784}0.46 &
				\cellcolor[HTML]{EEEEEE}0.007 &
				\cellcolor[HTML]{F0E784}0.46 &
				\cellcolor[HTML]{D4D4D4}0.013 &
				\cellcolor[HTML]{FFEB84}0.46 &
				\\
				&
				&
				&
				&
				&
				&
				&
				&
				&
				&
				\\
			\end{tabular}
		}
		\caption{Mean and standard deviation (Std) for the test weighted  $F_{1,w}$ score when ten times repeating experiments with varying number of random search iterations ($N_{\text{RS}}$) and ensemble size ($N_{\text{ens}}$) on six different datasets (Lipo, Desmoid, Liver, GIST, CRLM, and Melanoma). The color coding of the mean indicates the relative performance on each dataset (green: high; red: low); the color coding of the standard deviation indicates the relative variation on each dataset (dark: high; light: low).}
		\label{tab: rsens}
	\end{table*}
	
	\begin{figure*}
		\centering
		\includegraphics[width=0.975\textwidth]{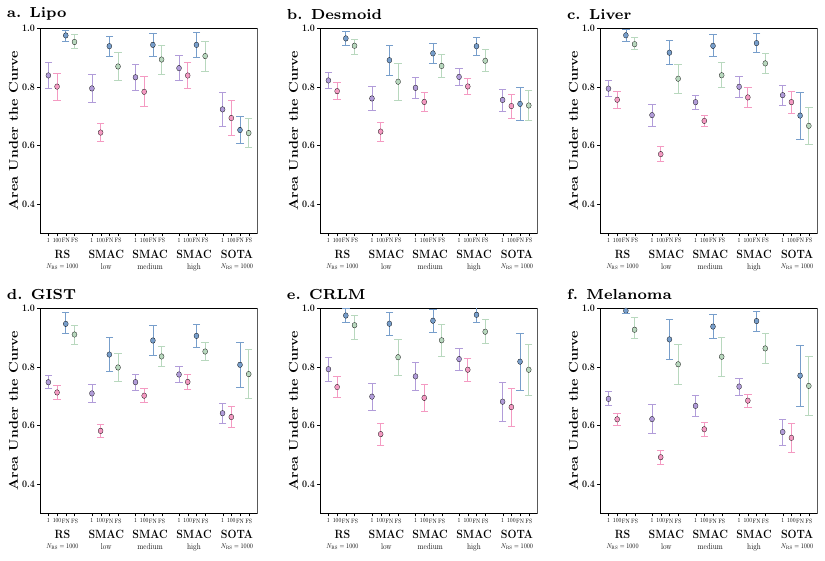}
		\caption{Error plots of the weighted $F_{1,w}$ score on the validation datasets of the radiomics models on six datasets (Lipo, Desmoid, Liver, GIST, CRLM, and Melanoma) for two optimization strategies (RS: random search, SMAC: sequential model-based algorithm configuration) with different computational budgets (low, medium, high) and a radiomics state-of-the-art (SOTA) baseline, when using either the single best found validation workflow (1) or one of three ensembling strategies (100: $N_{\text{ens}} = 100$, FN: fit number, FS: forward selection \cite{RN69}). The error plots represent the 95\% confidence intervals, estimated through $k_{\text{test}}=20$ random-split cross-validation on the validation dataset. The circle represents the mean.}
		\label{fig: errorplot_F1_validation}
	\end{figure*}
	
	
	\begin{table*}
		\centering
		\resizebox{0.9\textwidth}{!}{%
			\begin{tabular}{l | c c c c c c}
				\hline
				\textbf{Dataset} & \textbf{Lipo$^x$} & \textbf{Desmoid$^x$} & \textbf{Liver$^x$} & \textbf{GIST$^x$} & \textbf{CRLM$^x$} & \textbf{Melanoma$^x$} \\
				\hline
				AUC 			& 0.83 [0.76, 0.90] & 0.84 [0.77, 0.91] & 0.80 [0.73, 0.86] & 0.76 [0.70, 0.83] & 0.62 [0.49, 0.75] & 0.44 [0.32, 0.56] \\
				BCR 			& 0.74 [0.66, 0.82] & 0.75 [0.68, 0.82] & 0.73 [0.66, 0.80] & 0.71 [0.64, 0.77] & 0.60 [0.49, 0.71] & 0.46 [0.36, 0.55] \\
				$F_{1, w}$  	& 0.73 [0.65, 0.81] & 0.78 [0.72, 0.84] & 0.72 [0.66, 0.79] & 0.70 [0.64, 0.77] & 0.59 [0.48, 0.71] & 0.44 [0.33, 0.53] \\
				Sensitivity 	& 0.70 [0.57, 0.82] & 0.60 [0.47, 0.73] & 0.72 [0.61, 0.83] & 0.68 [0.57, 0.79] & 0.58 [0.40, 0.76] & 0.56 [0.39, 0.72] \\
				Specificity 	& 0.78 [0.67, 0.88] & 0.90 [0.84, 0.96] & 0.73 [0.63, 0.83] & 0.73 [0.65, 0.81] & 0.62 [0.43, 0.81] & 0.36 [0.18, 0.54] \\
				\hline
				\textbf{Dataset} & \textbf{HCC$^x$} & \textbf{MesFib$^x$} & \textbf{Prostate$^x$} & \textbf{Glioma$^b$} & \textbf{Alzheimer$^x$} & \textbf{H\&N$^x$} \\
				\hline
				AUC  			& 0.75 [0.67, 0.83] & 0.79 [0.68, 0.91] & 0.72 [0.61, 0.82] & 0.72 [0.62, 0.81] & 0.87 [0.84, 0.90] & 0.85 [0.79, 0.91] \\
				BCR  			& 0.69 [0.62, 0.77] & 0.70 [0.58, 0.81] & 0.67 [0.57, 0.78] & 0.62 [0.54, 0.70] & 0.78 [0.75, 0.81] & 0.75 [0.68, 0.83] \\
				$F_{1, w}$ 		& 0.69 [0.61, 0.77] & 0.69 [0.57, 0.81] & 0.67 [0.56, 0.78] & 0.55 [0.46, 0.65] & 0.80 [0.77, 0.83] & 0.76 [0.68, 0.83] \\
				Sensitivity 	& 0.72 [0.61, 0.84] & 0.74 [0.55, 0.92] & 0.67 [0.49, 0.85] & 0.42 [0.30, 0.53] & 0.69 [0.63, 0.75] & 0.82 [0.71, 0.93] \\
				Specificity 	& 0.66 [0.54, 0.79] & 0.66 [0.47, 0.85] & 0.68 [0.53, 0.82] & 0.82 [0.70, 0.94] & 0.87 [0.84, 0.90] & 0.69 [0.56, 0.83] \\
				\hline\\
			\end{tabular}
		}
		\caption{Classification results of our {\toolbox{WORC}} framework evaluated on the twelve datasets. For all metrics, the mean and 95\% confidence intervals (CIs) are reported. Abbreviations: AUC: area under the receiver operating characteristic curve; BCR: balanced classification rate \cite{RN704}; $F_{1, w}$: weighted F1-score. $^x$: 95\% CI constructed through a $k_{\text{test}}=100$ random-split cross-validation; $^b$: 95\% CI constructed through a 1000x bootstrap resampling of the test set.}
		\label{tab: allres}
	\end{table*}
	
	\begin{figure*}
		\centering
		\includegraphics[width=\textwidth]{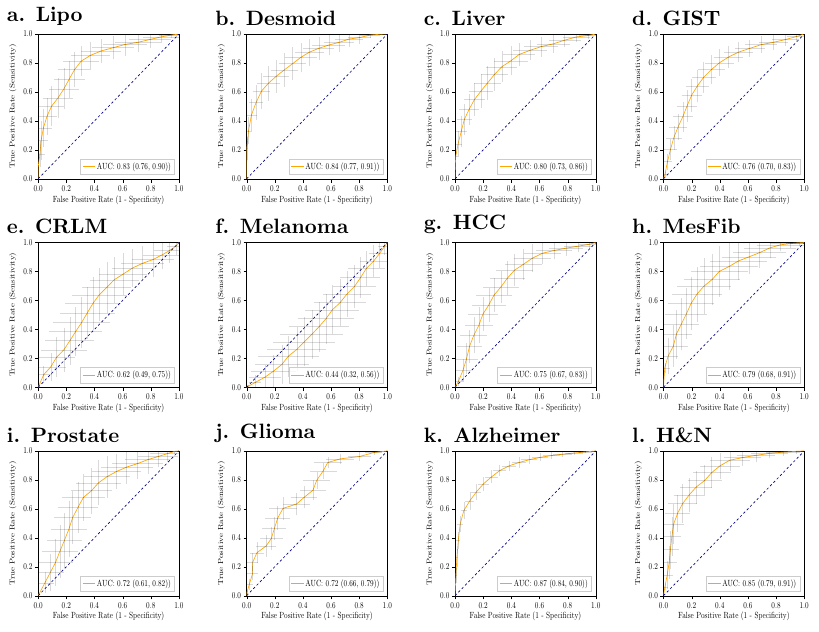}
		\caption{Receiver operating characteristic (ROC) curves of our {\toolbox{WORC}} framework evaluated on the twelve datasets. The crosses identify 95\% confidence intervals of $k_{\text{test}}=100$ random-split cross-validation with the orange curve fit through the means, except for the glioma dataset, where 95\% confidence intervals are constructed through a 1000x bootstrap resampling of the test set with the orange curve representing the point estimate. The numbers in the legend represent the mean (95\% confidence interval) of the area under the curve (AUC).} 
		\label{fig: ROC}
	\end{figure*}
	
	\begin{table*}
		\centering
			\begin{tabular}{l | c c l l }
				\hline
				\textbf{Dataset} & \textbf{WORC} & \textbf{Mean of clinicians } & \textbf{Scores of clinicians} & \textbf{Reference}\\
				\hline
				Lipo$^x$ & 0.83 & 0.69 & 0.74, 0.72, and 0.61 & \cite{RN574} \\
				Desmoid$^{*,x}$ & 0.84 & 0.84 & 0.88, and 0.80 & \cite{RN678}\\
				Liver$^{*,x}$ & 0.80 & 0.85 & 0.86 and 0.83 & \cite{starmans2024BLT} \\
				GIST$^x$ & 0.76 & 0.76 & 0.84, 0.76, and 0.69 & \cite{RN1246} \\
				MesFib & 0.79 & 0.75 & 0.85, 0.82, 0.76, 0.71, and 0.60 & \cite{RN923} \\
				Prostate$^*$ & 0.72 & 0.47 & 0.50 and 0.44 & \cite{RN868} \\
				Glioma & 0.72 & 0.66 & 0.83, 0.79, 0.58 and 0.45 & \cite{RN492} \\
				\hline\\
			\end{tabular}
		\caption{Performance of our {\toolbox WORC} radiomics framework compared to visual scoring by several clinicians. For the clinicians, both the mean of all clinicians and the separate scores are provided. $^*$: the visual scoring by the clinicians was evaluated on a different dataset than the {\toolbox WORC} framework or on a subset of the data. $^x$: the clinicians were given additional information (e.g.  age, sex) besides the imaging, while the {\toolbox WORC} framework was only given the images. Abbreviations: AUC: area under the receiver operating characteristic curve.}
		\label{tab: ScoreClinicians}
	\end{table*}
	
	\begin{table*}
		\resizebox{\textwidth}{!}{%
			\begin{tabular}{llllllllll}
				\toprule
				\textbf{\begin{tabular}[c]{@{}l@{}}Histogram \\ (13 features)\end{tabular}}                                                                                                                                                                                                                                                                                & \multicolumn{2}{l}{\textbf{\begin{tabular}[c]{@{}l@{}}LoG \\ (13*3=39 features)\end{tabular}}}                                                                                                                                                                                                                                                                                                                                                       & \textbf{\begin{tabular}[c]{@{}l@{}}Vessel \\ (12*3=39 features)\end{tabular}}                                                                                                                                                                                                                           & \multicolumn{2}{l}{\textbf{\begin{tabular}[c]{@{}l@{}}GLCM (MS)\\ (6*3*4*2=144 features)\end{tabular}}}                                                                                                                                                                                                                                                                                                                                                                                                                                                                                                                                                          & \textbf{\begin{tabular}[c]{@{}l@{}}Gabor \\ (13*4*3=156 features)\end{tabular}}                                                                                                                                                                                                                                                                                                                                                                                                                                                                                                                                                                                               & \textbf{\begin{tabular}[c]{@{}l@{}}NGTDM \\ (5 features)\end{tabular}}                                                                                                             & \textbf{\begin{tabular}[c]{@{}l@{}}LBP \\ (13*3=39 features)\end{tabular}}                                                                                                                                                         &  \\
				\midrule
				\begin{tabular}[c]{@{}l@{}}min\\ max\\ mean\\ median\\ std\\ skewness\\ kurtosis\\ peak\\ peak position\\ range\\ energy\\ quartile range\\ entropy\end{tabular}                                                                                                                         & \multicolumn{2}{l}{\begin{tabular}[c]{@{}l@{}}min\\ max\\ mean\\ median\\ std\\ skewness\\ kurtosis\\ peak\\ peak position\\ range\\ energy\\ quartile\\ entropy\end{tabular}}                                                                                                                                                                                                     & \begin{tabular}[c]{@{}l@{}}min\\ max\\ mean\\ median\\ std\\ skewness\\ kurtosis\\ peak\\ peak position\\ range\\ energy\\ quartile\\ entropy\end{tabular}                                                                            & \multicolumn{2}{l}{\begin{tabular}[c]{@{}l@{}}contrast (normal, MS mean +   std) \\ dissimilarity (normal, MS   mean + std)\\ homogeneity (normal, MS mean   + std)\\ angular second moment (ASM)   (normal, MS mean + std)\\ energy (normal, MS mean +   std)\\ correlation (normal, MS mean   + std) \\ \\ \\ \\ \\ \\ \\ \\ \end{tabular}}                                                                                                                                                                                                                                                                                                                     & \begin{tabular}[c]{@{}l@{}}min\\ max\\ mean\\ median\\ std\\ skewness\\ kurtosis\\ peak\\ peak position\\ range\\ energy\\ quartile range\\ entropy\end{tabular}                                                                                                                                                                                                                                                                                                                                                                                                                                            & \begin{tabular}[c]{@{}l@{}}busyness \\ coarseness\\ complexity\\ contrast\\ strength \\ \\ \\ \\ \\ \\ \\ \\ \\ \end{tabular}                                                                & \begin{tabular}[c]{@{}l@{}}min\\ max\\ mean\\ median\\ std\\ skewness\\ kurtosis\\ peak\\ peak position\\ range\\ energy\\ quartile range\\ entropy\end{tabular} &  \\
				\midrule
				\multicolumn{2}{l}{\textbf{\begin{tabular}[c]{@{}l@{}}GLSZM \\ (16 features)\end{tabular}}}                                                                                                                                                                                                                                                                                                                                                                                                                                                                                                                     & \multicolumn{3}{l}{\textbf{\begin{tabular}[c]{@{}l@{}}GLRM \\ (16 features)\end{tabular}}}                                                                                                                                                                                                                                                                                                                                                                                                                                                                                                    & \textbf{\begin{tabular}[c]{@{}l@{}}GLDM\\ (14 features)\end{tabular}}                                                                                                                                                                                                                                                                                                                                                                                                                                                                                                              & \textbf{\begin{tabular}[c]{@{}l@{}}Shape \\ (35 features)\end{tabular}}                                                                                                                                                                                                                                                                                                                                                                                                                                                                                                                                                                                                       & \textbf{\begin{tabular}[c]{@{}l@{}}Orientation \\ (9 features)\end{tabular}}                                                                                                       & \textbf{\begin{tabular}[c]{@{}l@{}}Local phase \\ (13*3=39 features)\end{tabular}}                                                                                                                                                 &  \\
				\midrule
				\multicolumn{2}{l}{\begin{tabular}[c]{@{}l@{}}Gray Level Non   Uniformity\\ Gray Level Non   Uniformity Normalized\\ Gray Level Variance\\ High Gray Level Zone   Emphasis\\ Large Area Emphasis\\ Large Area High Gray   Level Emphasis\\ Large Area Low Gray Level   Emphasis\\ Low Gray Level Zone   Emphasis\\ SizeZoneNonUniformity\\ SizeZoneNonUniformityNormalized\\ SmallAreaEmphasis\\ SmallAreaHighGrayLevelEmphasis\\ SmallAreaLowGrayLevelEmphasis\\ ZoneEntropy\\ ZonePercentage\\ ZoneVariance \\ \\ \\ \\ \\ \end{tabular}} & \multicolumn{3}{l}{\begin{tabular}[c]{@{}l@{}}Gray Level Non Uniformity\\ Gray Level Non Uniformity   Normalized\\ Gray Level Variance\\ High Gray Level Run Emphasis\\ Long Run Emphasis\\ Long Run High Gray Level   Emphasis\\ Long Run Low Gray Level   Emphasis\\ Low Gray Level Run Emphasis\\ RunEntropy\\ RunLengthNonUniformity\\ RunLengthNonUniformityNormalized\\ RunPercentage\\ RunVariance\\ ShortRunEmphasis\\ ShortRunHighGrayLevelEmphasis\\ ShortRunLowGrayLevelEmphasis \\ \\ \\ \\ \\ \end{tabular}} & \begin{tabular}[c]{@{}l@{}}Dependence Entropy\\ Dependence Non-Uniformity\\ Dependence Non-Uniformity   Normalized\\ Dependence Variance\\ Gray Level Non-Uniformity\\ Gray Level Variance\\ High Gray Level Emphasis\\ Large Dependence Emphasis\\ Large Dependence High Gray   Level Emphasis\\ Large Dependence Low Gray   Level Emphasis\\ Low Gray Level Emphasis\\ Small Dependence Emphasis\\ Small Dependence High Gray   Level Emphasis\\ Small Dependence Low Gray   Level Emphasis \\ \\ \\ \\ \\ \\ \\ \end{tabular} & \begin{tabular}[c]{@{}l@{}}compactness (mean + std)\\ radial distance (mean + std)\\ roughness (mean + std) \\ convexity (mean + std) \\ circular variance (mean +   std)\\ principal axes ratio (mean +   std)\\ elliptic variance (mean +   std) \\ solidity (mean + std)\\ area (mean, std, min + max\\ volume (total, mesh, volume)\\ elongation\\ flatness\\ least axis length\\ major axis length\\ minor axis length\\ maximum diameter 3D\\ maximum diameter 2D (rows,   columns, slices)\\ sphericity\\ surface area\\ surface volume ratio\end{tabular} & \begin{tabular}[c]{@{}l@{}}theta\_x\\ theta\_y\\ theta\_z\\ COM index x\\ COM index y\\ COM index z\\ COM x\\ COM y\\ COM z \\ \\ \\ \\ \\ \\ \\ \\ \\ \\ \\ \\ \end{tabular} & \begin{tabular}[c]{@{}l@{}}min\\ max\\ mean\\ median\\ std\\ skewness\\ kurtosis\\ peak\\ peak position\\ range\\ energy\\ quartile\\ entropy \\ \\ \\ \\ \\ \\ \\ \\ \end{tabular}       & \\
				\bottomrule
				\multicolumn{9}{p{1.44\textwidth}}{*Abbreviations: COM: center of mass; GLCM: gray level co-occurrence matrix; MS: multi slice; NGTDM: neighborhood gray tone difference matrix; GLSZM: gray level size zone matrix; GLRLM: gray level run length matrix; LBP: local binary patterns; LoG: Laplacian of Gaussian; std: standard deviation.}
			\end{tabular}%
		}
		\caption{Overview of the 564 features used by default in the {\toolbox{WORC}} framework. GLCM features were calculated in four different directions (0, 45, 90, 135 degrees) using 16 gray levels and pixel distances of 1 and 3. LBP features were calculated using the following three parameter combinations: 1 pixel radius and 8 neighbors, 2 pixel radius and 12 neighbors, and 3 pixel radius and 16 neighbors. Gabor features were calculated using three different frequencies (0.05, 0.2, 0.5) and four different angles (0, 45, 90, 135 degrees). LoG features were calculated using three different widths of the Gaussian (1, 5 and 10 pixels). Vessel features were calculated using the full mask, the edge, and the inner region. Local phase features were calculated on the monogenic phase, phase congruency and phase symmetry.}
		\label{tab: Features}
	\end{table*}
	
	\begin{algorithm*}
		\caption{Minimal working example of the {\toolbox{WORC}} toolbox interface in Python}\label{alg: WORCCodeExample}
\begin{lstlisting}[language=Python, basicstyle=\footnotesize]
from WORC import SimpleWORC

# Create a Simple WORC object
experiment = SimpleWORC("experiment_name")

# Set the input data according to the variables we defined earlier
experiment.images_from_this_directory(
    directory='/datadir',
    image_file_name='image.nii.gz')
experiment.segmentations_from_this_directory(
    directory='/datadir',
    segmentation_file_name='mask.nii.gz')
experiment.labels_from_this_file(label_file='labels.xlsx')
experiment.predict_labels(label_name='diagnosis')

# Set the types of images WORC has to process. Used in fingerprinting
experiment.set_image_types(['CT'])

# Use the standard workflow for binary classification
experiment.binary_classification(coarse=False)

# Optional: change a configuration field to only use SVMs for classification
overrides = {
            'Classification': {
                'classifiers': 'SVM',
                },
            }
experiment.add_config_overrides(overrides)

# Run the experiment!
experiment.execute()
\end{lstlisting}
	\end{algorithm*}
	
}

\vfill

\end{document}